\documentclass[iop]{emulateapj-rtx4} 
\shortauthors{Sekanina}
\shorttitle{Kreutz Sungrazers:\ Recent Modeling and the SOHO Objects}
\slugcomment{Version \today}
\newcommand{\Rsun}{$R_{\mbox{\scriptsize \boldmath $\odot$}}$}

\begin{document}
\title{KREUTZ SUNGRAZERS:\ SUMMARY OF RECENT MODELING AND ORBITS OF THE SOHO OBJECTS}
\author{Zdenek Sekanina}
\affil{Jet Propulsion Laboratory, California Institute of Technology,
  4800 Oak Grove Drive, Pasadena, CA 91109, U.S.A.}
\email{Zdenek.Sekanina@jpl.nasa.gov.}

\begin{abstract} 
I summarize and streamline the results of recent modeling of the orbital evolution  
and cascading fragmentation of the Kreutz sungrazers.  The model starts with        
Aristotle's comet --- the progenitor whose nucleus is assumed to be a contact       
binary --- splitting near aphelion into the two lobes and concludes with the        
SOHO dwarf objects as the end products of the fragmentation process.                
The Great March Comet of 1843, a member of Population~I, and the Great September
Comet~of~1882, a member of Population~II, are deemed the largest surviving
masses of the lobes.  I establish that the Kreutz system consists currently of
nine populations, one of which --- associated with comet \mbox{Pereyra ---}
is a side branch of Population~I.  The additions to the Kreutz system
proposed as part of the new model are the daylight comets of AD~363,
recorded by the Roman historian Ammianus Marcellinus, and the Chinese
comets of September 1041 and September 1138, both listed in Ho's
catalogue.  The comets of 363 are the first-generation fragments, the latter
--- together with the Great Comet of 1106 --- the second-generation fragments.
Attention is directed toward the populations' histograms of perihelion              
distance of the SOHO sungrazers and the plots of this distance as a                 
function of the longitude~of~the ascending node.  Arrival of bright, naked-eye      
Kreutz sungrazers in the coming decades is predicted.                               
\end{abstract} 
\section{Introduction}  
Recent introduction of a new model for the orbital evolution of the Kreutz
system of sungrazing comets (Seka\-nina 2021a; hereafter referred to as Paper~1)
was necessitated by significant developments in the past 15~years, which made
major parts of the previous models, including the two-superfragment model
(Sekanina 2002; Seka\-nina \& Chodas 2004), obsolete.  Among the developments,
the most important event was the arrival of the naked-eye sungrazer Lovejoy
(C/2011~W3), whose orbit defied Marsden's (1989) expanded classification of
the Kreutz comets into three populations --- or subgroups in his terminology
--- I, II, and IIa.  Sekanina \& Chodas (2012) referred to comet Lovejoy as
a member of a new population --- III, a designation adopted in Paper~1.

The new model is based on two stipulations, not employed in the earlier models.
One is the progenitor in shape of a contact binary, consisting of two lobes of
comparable dimensions connected with a neck.  The progenitor's splitting into
the separate lobes (and, simultaneously or subsequently, the neck) is much more
plausible than the breakup of a nearly spherical nucleus into two halves through
its center, the scenario tolerated in the two-superfragment model (Sekanina \&
Chodas 2004); the contact binary also emerges as an apparently rather common
figure among cometary nuclei and Kuiper Belt objects (e.g., Sierks et al.\ 2015;
Stern et al.\ 2019).

My second stipulation was that most of the surviving mass of Lobe~I was contained
in the Great March Comet of 1843 or C/1843~D1, the intrinsically brightest known
fragment of Population~I, and most of the surviving mass of Lobe~II in the Great
September Comet of 1882 or C/1882~R1, intrinsically the brightest known fragment
of Population~II.  The orbital distance that separated the two masses, gradually
increasing with time, was in the 19th century equivalent to a time difference of
merely 39.5~yr in perihelion time, a measure of the young age of the Kreutz system.
This constraint effectively eliminated all scenarios that involved 5th~century
comets as members of the Kreutz system because they would have implied orbital
periods much too short to fit the arrival times of the two 19th~century's
spectacular sungrazers. 

Another recent development was the detection in Paper~1 of as many as nine
populations of sungrazers that are making up the Kreutz system.  This major result
was obtained by processing Marsden's gravitational orbital elements of the SOHO
Kreutz sungrazers (Section~2) and is understood in the context of cascading
fragmentation:\ the initial breakup was followed by a number of events of
secondary splitting.  The account and results of the new model are below
summarized and streamlined with the aim to benefit the reader.

The problem of systematic trends in the gravitational orbital elements of the
SOHO sungrazers was explored in great detail on several particular objects by
Sekanina \& Kracht (2015).  The most troublesome were the systematic positional
deviations of the SOHO sungrazers' lines of apsides from the fixed position of
the line of apsides of the naked-eye Kreutz comets.  The deviations, apparent
in the plots of the inclination and perihelion latitude against the longitude of
the ascending node, were shown to be effects of the normal component of the
nongravitational acceleration, while a set of select data of higher accuracy
in Paper~1 revealed the structure of the system of SOHO Kreutz sungrazers to be
much more complex than earlier thought.  A further, not yet fully examined test,
a potential relationship between the longitude of the ascending node and the
perihelion distance, is a subject of the present paper.

\begin{figure*}
\vspace{-5.68cm}
\hspace{-1.5cm}
\centerline{
\scalebox{0.75}{
\includegraphics{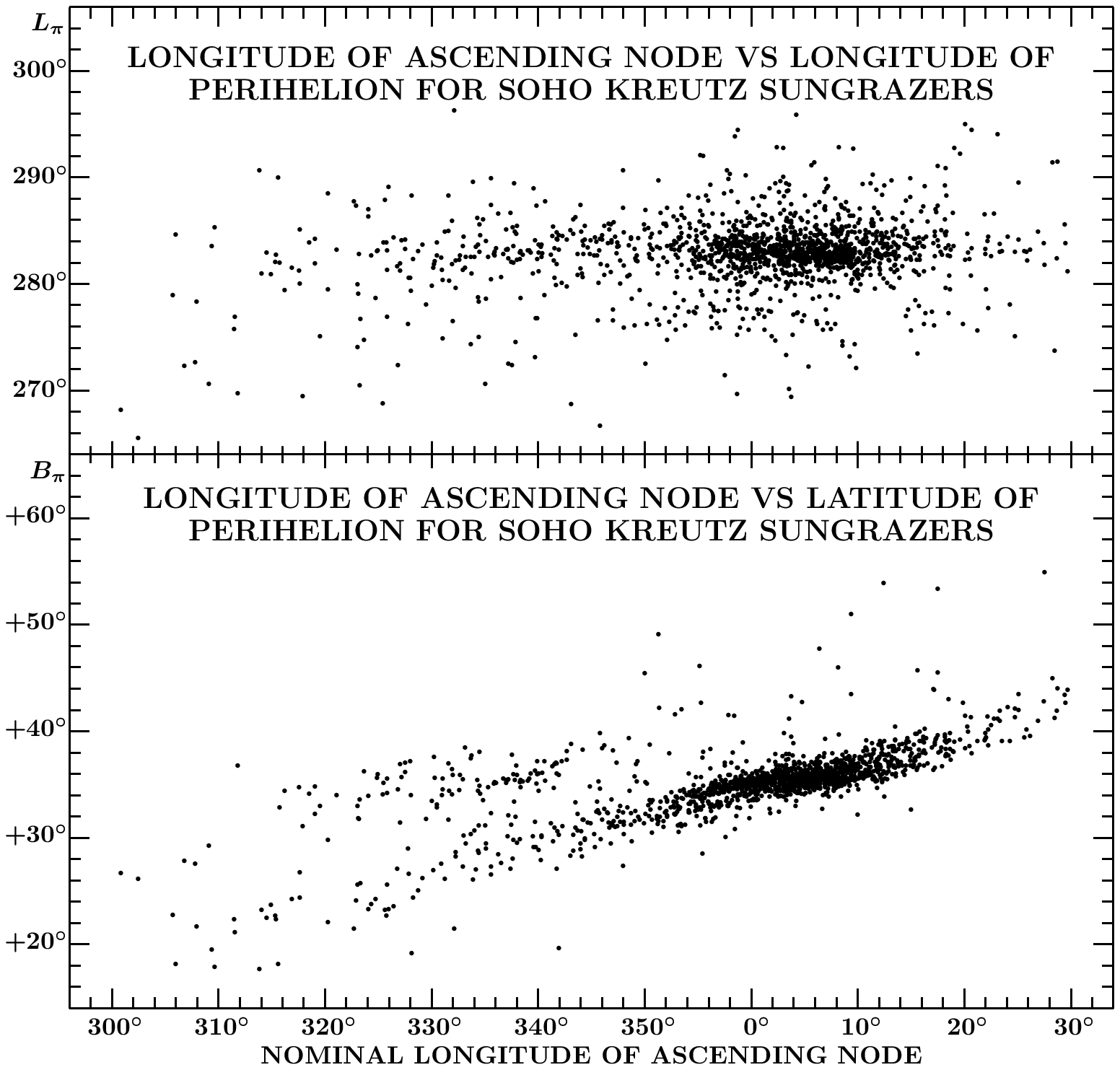}}} 
\vspace{-5.25cm}
\caption{Plot of the nominal perihelion longitude $L_\pi$ (at the top) and latitude
 $B_\pi$ against the nominal longitude of the ascending node $\Omega$ for 1565~SOHO
 Kreutz sungrazers that arrived between January 1996 and June 2010.  While $L_\pi$
 stays, by and large, constant, $B_\pi$ increases systematically over an interval
 of nearly 90$^\circ$ in $\Omega$. (Reproduced from Sekanina \& Kracht
 2015.){\vspace{0.78cm}}}
\end{figure*}

\section{Nominal Orbital Elements of the SOHO\\Dwarf Sungrazers} 
Marsden computed sets of approximate parabolic gravitational orbital elements for
more than 1500 Kreutz sungrazers detected in the coronagraphic images of the SOHO
space observatory between 1996 and 2010.  The orbits are mostly in Marsden \&
Williams (2008), for the objects that arrived during 2008--2010 in a number of
{\it Minor Planet Circular\/} issues.  I refer to these sets of elements as {\it
nominal\/} to distinguish them from the sets of {\it true\/} elements.  The
difference is significant, as the SOHO sungrazers were subjected to an appreciable
sublimation-driven force ignored by Marsden.  The magnitude of this acceleration,
pointing --- as already noted --- essentially in the direction normal to the
orbital plane, was in extreme cases nearly comparable to the Sun's gravitational
acceleration.  The computation of the true orbit was time consuming, because the
parameter $A_3$ of the normal component of the nongravitational acceleration has
to be determined iteratively by forcing the standard position of the line of
apsides, as illustrated by Sekanina \& Kracht (2015) for several particular SOHO
Kreutz sungrazers.

It was inconceivable to compute the true orbits for all 1500 SOHO comets.  Instead,
I chose to exploit the implications of the dependence of the nominal perihelion
latitude, $B_\pi$, on the nominal longitude of the ascending node, $\Omega$,
presented for the 1500 SOHO Kreutz sungrazers in Figure~1.  The reason for the
great deal of noise that the figure exhibits is Marsden's computation of the
nominal orbits from astrometric positions derived from measurements of {\it all\/}
images, taken with the C2 as well as C3 coronagraphs on board the SOHO spacecraft.
Because the pixel size of the C3 coronagraph is about five times larger than the
pixel size of the C2 coronagraph, an obvious way to improve the quality of the
$B_\pi(\Omega)$ plot was to restrict the data set to the objects whose nominal
orbits were computed exclusively from the C2 images.  The 193~select SOHO
sungrazers are listed in Table~1 and the plot of their latitude $B_\pi$ against
nodal longitude $\Omega$ is presented in Figure~2.

\begin{table}[t]
\vspace{-2.5cm}
\hspace{0.4cm}
\centerline{
\scalebox{0.812}{
\includegraphics{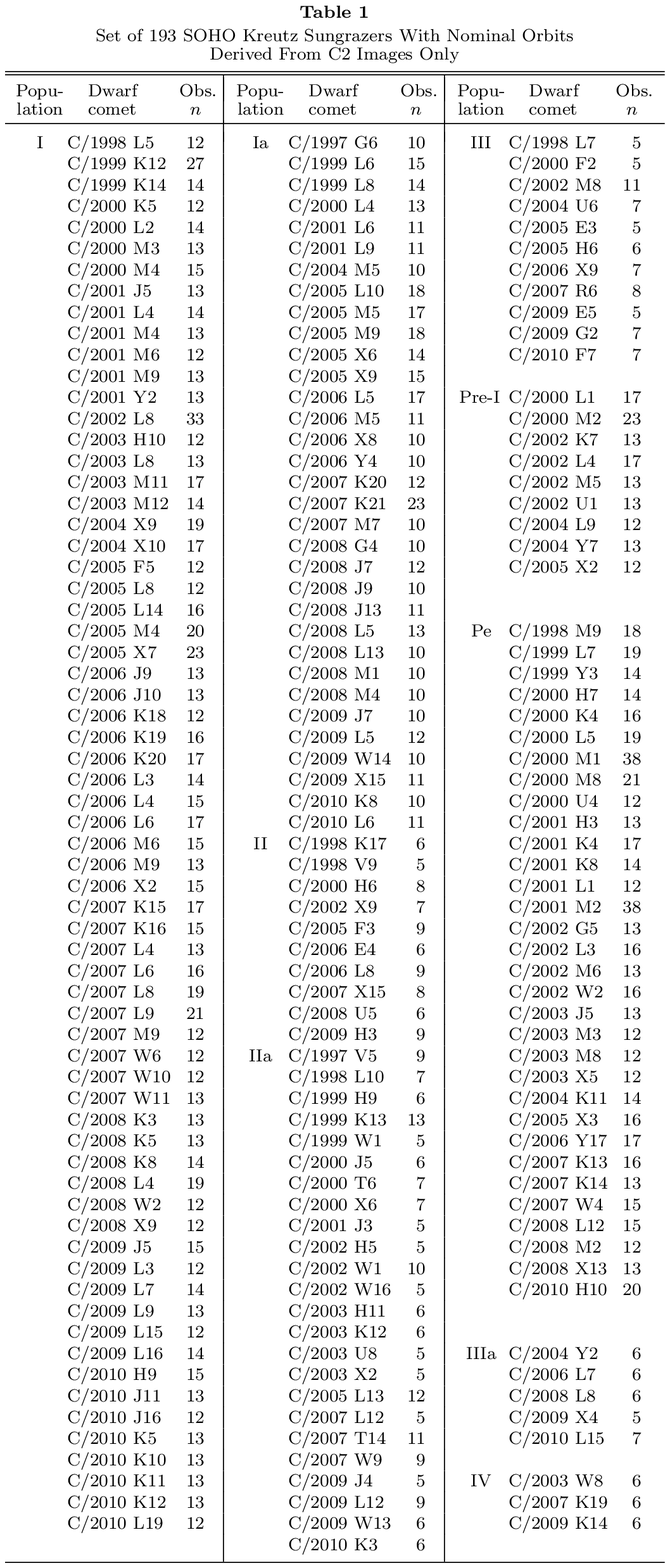}}} 
\vspace{-0.8cm}
\end{table}

It is well known that an overwhelming majority of the SOHO dwarf sungrazers
belongs to Population~I.  This turns out as well to be the case for the subset
of these objects with the orbits derived exclusively from the data based on
the C2 coronagraphic images.  Since the purpose of the exercise in Paper~1 was
to obtain a data set of the highest possible quality, I included only those
sungrazers of Population~I whose orbits were derived from at least 12~astrometric
observations.  As I was unaware of Populations~Pe and Pre-I at the time of this
selection process, their sets were subjected to the same constraint.  On the other
hand, I did suspect the existence of Population~Ia, and with the aim to increase
the number of its members in the plot, I relaxed the constraint for this population
from the minimum of 12~observations down to 10.  The numbers of members in all
the other populations in Table~1 were considerably lower and the minimum of
observations for them was further relaxed to five; I deemed all orbits based
on fewer than five observations too uncertain to include in the data set.  

The plot of the 193 nominal values of $B_\pi$ as a function of $\Omega$ in Figure~2,
reproduced from Paper~1, shows that the data points line up along the straight lines
\begin{equation}
B_\pi = \widehat{B}_\pi + b \left( \Omega \!-\! \widehat{\Omega} \right), 
\end{equation}
where $\widehat{B}_\pi$ is the constant true perihelion latitude, $\widehat{\Omega}$
is the true longitude of the ascending node for the given population of sungrazers,
and $b$ is{\vspace{-0.07cm}} the slope, which equals \mbox{$b = 0.28$}.  Equation~(1)
serves to compute $\widehat{\Omega}$ from the nominal values of $\Omega_k$ and
$(B_\pi)_k$ for $n$ SOHO sungrazers:
\begin{equation}
\widehat{\Omega} = \frac{\widehat{B}_\pi}{b} + \frac{1}{n} \sum_{k=1}^{n} \left[
 \Omega_k - \frac{1}{b} (B_\pi)_k \right]. 
\end{equation}
The values of the other{\vspace{-0.07cm}} true angular elements --- the argument of
perihelion, $\widehat{\omega}$, and{\vspace{-0.07cm}} inclination, $\widehat{i}$,
follow from the values of $\widehat{\Omega}$, $\widehat{B}_\pi$, and the true
perihelion{\vspace{-0.06cm}} longitude, $\widehat{L}_\pi$:
\begin{eqnarray}
\cos \widehat{\omega} & = & \cos \widehat{B}_\pi \cos (\widehat{L}_\pi \!-\!
 \widehat{\Omega}), \nonumber \\
\tan \widehat{i} & = & \tan \widehat{B}_\pi \csc (\widehat{L}_\pi \!-\!
 \widehat{\Omega}), 
\end{eqnarray}
where for the{\vspace{-0.07cm}} Kreutz sungrazers \mbox{$\widehat{L}_\pi \simeq
283^\circ$}, \mbox{$\widehat{B}_\pi \simeq +35^\circ$}, \mbox{$\widehat{i} >
90^\circ$}, and \mbox{$0^\circ \!< \widehat{\omega} < 90^\circ$}.  These
and all other angular elements are in this paper referred to the J2000 equinox.

It was the plot presented here in Figure~2, on which the existence of the nine
populations of the Kreutz system was first recognized in Paper~1.  Added to
the already known Populations I, II, IIa, and III were --- in the order of the
decreasing nodal longitude --- Pre-I, Pe (apparently a branch of Population~I
associated with comet Pereyra, C/1963~R1, rather than with the Great March
Comet of 1843), Ia, IIIa, and IV.

In order to improve clarity of Figure~2, the crowded areas around Population~I
are redrawn in greater detail in Figure~3.  The sets of true orbital elements
averaged over all members from Table~1 are listed for all nine populations in
the columns {\it dwarfs\/} in Table~2, also reproduced from Paper~1.  Where
available, the orbit of the associated brightest known naked-eye Kreutz
sungrazer is listed in the column {\it main\/}.  For Population~Ia the orbit
shown in the column {\it model\/} refers to the progenitor, as derived in
Paper~1. 

The remarkable feature of Table~2 is a relatively uniform step in the true
longitude of the ascending node between the neighboring populations.  The
average size of this step, near 10$^\circ$ (except for Population~Pe), is
significantly greater than the minor differences between the columns {\it
main\/} (or {\it model\/}) and {\it dwarfs\/}.  From the SOHO dwarf
sungrazers of Table~1, supported by the naked-eye, Solwind (P78-1), and
Solar Maximum Mission (SMM) objects, strong evidence has been assembled
to support a new investigation of the process of cascading fragmentation
of the Kreutz system's progenitor.

\begin{figure*}[ht]
\vspace{-0.45cm}
\hspace{-0.25cm}
\centerline{
\scalebox{0.79}{
\includegraphics{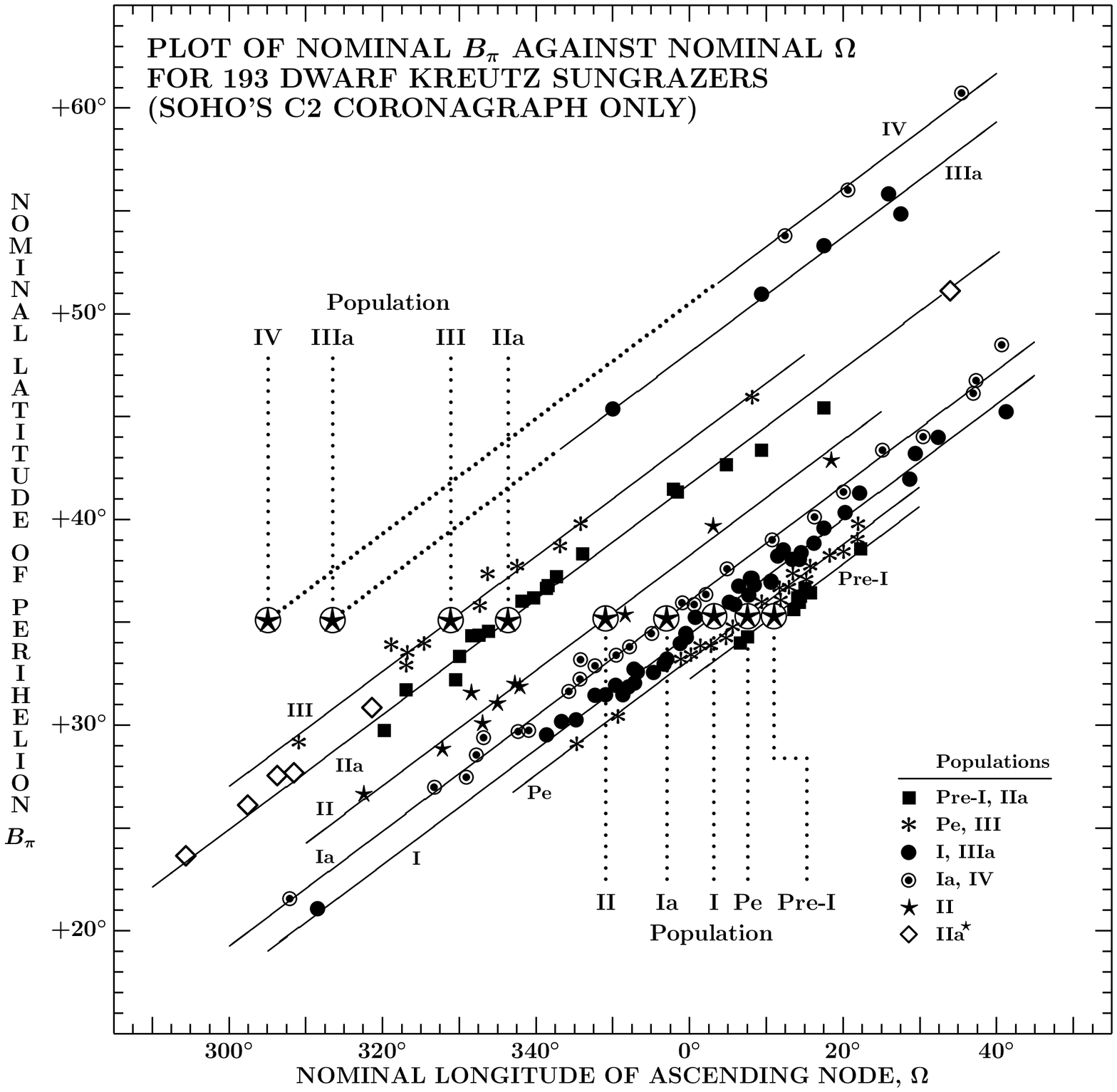}}} 
\vspace{-7.3cm}
\caption{Plot of the nominal latitude of perihelion, $B_\pi$, as a function of the
nominal longitude of the ascending node, $\Omega$ (equinox J2000), for the 193 dwarf
Kreutz sungrazers imaged in 1997--2010 exclusively with the C2 coronagraph on board
the SOHO Space Observatory; their gravitational orbits were computed by Marsden.  The
data points cluster fairly tightly along a set of straight lines of a constant slope of
\mbox{$dB_\pi/d\Omega = +0.28$} and refer to one of {\it nine\/} sungrazer populations.
Each line crosses the standard perihelion latitude of the population (Table~2) at a
point whose abscissa determines the respective population's {\it true\/} nodal longitude
(i.e., corrected for effects of the nongravitational acceleration).  In the order of
decreasing true nodal longitude, the populations are Pre-I, Pe, I, Ia, II, IIa, III,
IIIa, and IV, as depicted in the plot; Population~Pe is a side branch of Population~I.
All members of a population are plotted with the same symbol; the
exception is Population IIa, whose members with anomalously small perihelion distances,
IIa{\boldmath $^{\!\star}\!$}, are identified by symbols that differ from the symbols
for the remaining members.  The position of the {\it true\/} longitude of the ascending
node is for each population highlighted by an oversized circled star.  Some data points,
which were overlapping any of these major symbols or were contributing to one of
awkward-looking local clumps, were for the sake of clarity either removed or
slightly shifted along the slope of the fitting lines. (Reproduced from
Paper~1.){\vspace{0.7cm}}}
\end{figure*}

\section{Nuclear Splitting:\ Separation Velocities and Perturbed Orbits of
 Fragments.\\Populations and Clusters} 
Cometary nuclei fragment by a variety of forces.  Sungrazers split at or near
perihelion by the Sun's tidal force.  This mechanism was in the past so popular
that until recently all sungrazers were thought to break up exclusively in the
immediate proximity of the Sun, in spite of the fact that efforts to explain
the orbital evolution of the Kreutz system in this fashion met with formidable
difficulties and were never successful.

%
\begin{figure*}
\vspace{-6.6cm}
\hspace{-1.2cm}
\centerline{
\scalebox{0.76}{
\includegraphics{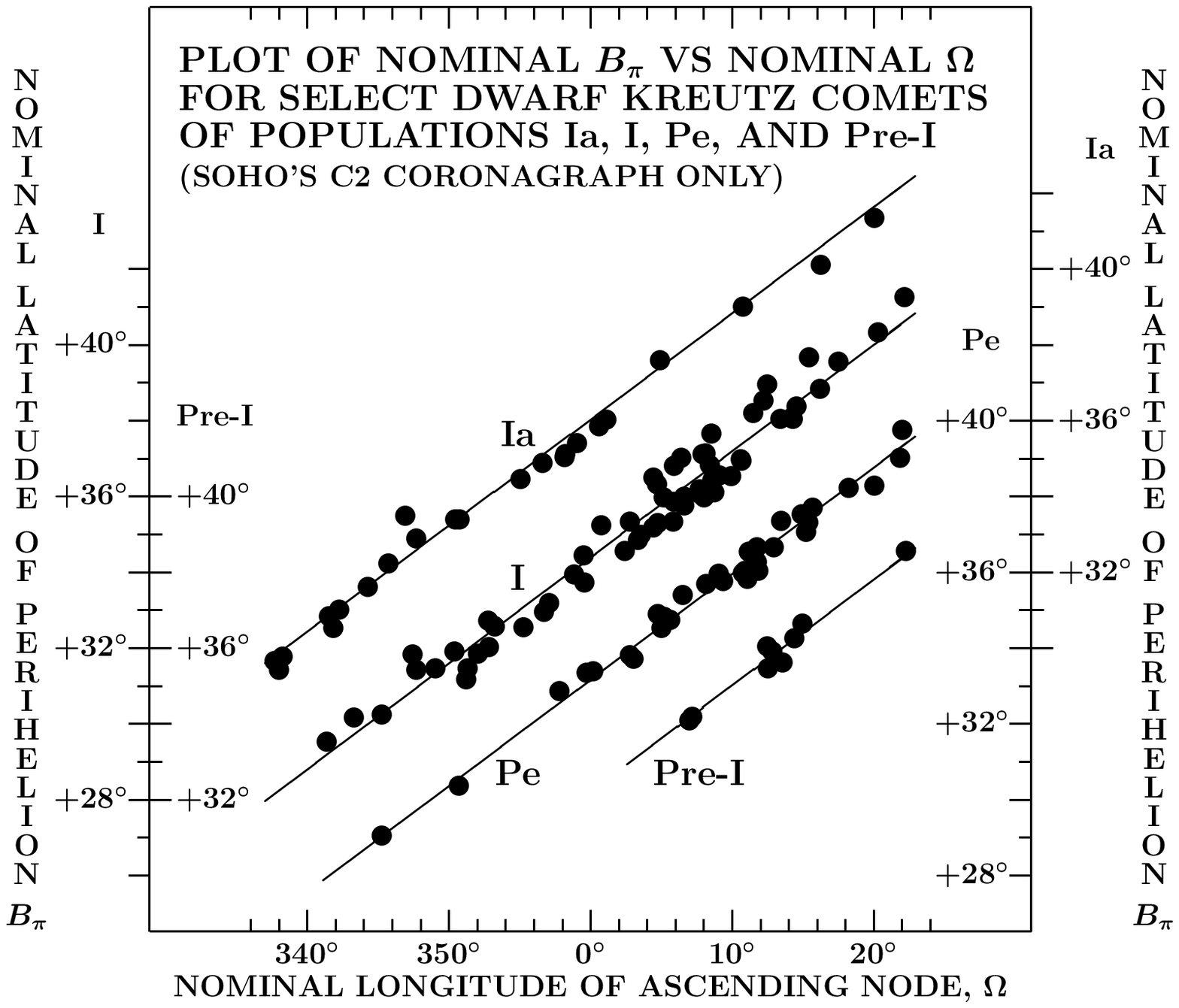}}} 
\vspace{-5.31cm}
\caption{Detail of the plot of the nominal latitude of perihelion, $B_\pi$,
against the nominal longitude of the ascending node, $\Omega$ (equinox J2000),
for segments of Populations~I, Ia, Pre-I, and Pe with the ordinate scales
systematically shifted for the sake of clarity.{\vspace{0.9cm}}}
\end{figure*}

On the other hand, it is well known that comets fairly often split far from the
Sun and planets, and there is no reason why the sungrazers should do otherwise.
Some 20 years ago I proposed (Sekanina 2002) that a Kreutz comet can fragment
at any point of its orbit about the Sun and that the fragmentation process
has cascading nature.  This is of key importance because it turns out that the
new orbit, in which a fragment ends up after breaking off from its sungrazing
parent, depends dramatically on the orbital location of the event.  When
the breakup occurs at or near perihelion, the fragment's separation velocity
--- possibly of rotational origin and typically not exceeding a few meters
per second --- affects almost exclusively the period of the orbit, the
difference compared to the parent's orbit reaching as much as several
hundred years; the aphelion distance does of course change accordingly.
On the other hand, when the breakup takes place near aphelion (say, about
160~AU from the Sun), significantly affected are (i)~the angular elements by
the normal component of the separation velocity, shifting by $\sim$10$^\circ$
or more in the nodal longitude; and (ii)~the perihelion distance by the
transverse component, changing by a few tenths of the solar radius, so that
the fragment's perihelion point may get below the surface of the photosphere.

Conversely, it is impossible for a Kreutz fragment to end up in an orbit whose
period differs from the orbit of its parent by more than a few months following
an event occurring at or close to aphelion; and it is equally impossible
for a fragment to end up in an orbit whose angular elements and perihelion
distance differ from those of its parent by any nontrivial amounts in an
event taking place at or close to perihelion.

Fragments generated by tidal disruption of a sungrazer are expected to end up
in orbits with dramatically different periods even if they should acquire no
differential momenta at breakup.  This is so because, in general, the centers of
mass of the fragments are at the time of their birth located at slightly different
heliocentric distances (Figure~4) but have the same orbital velocity:\ the fragment
farther from the Sun has to get into a greater orbit, with a longer period.  If
the heliocentric distance of the parent sungrazer at the time of fragmentation is
$r_{\rm frg}$ (at or near perihelion) and the difference in the radial distances
between a fragment and the parent is $\Delta U_{{\rm p}\rightarrow{\rm f}}(r_{\rm
frg},P_{\rm par})$, the fragment's orbital period, $P_{\rm frg}$, is related to
the parent's orbital period, $P_{\rm par}$, by (Sekanina \& Kracht 2022; hereafter
referred to as Paper~2)
\begin{equation}
P_{\rm frg} = P_{\rm par} \! \left[ 1 - \frac{2 \Delta U_{{\rm p}\rightarrow{\rm f}}
 (r_{\rm frg}, P_{\rm par})}{r_{\rm frg}^2} P_{\rm par}^{\frac{2}{3}}
 \right]^{-\frac{3}{2}} \!\!\!, 
\end{equation}
where the orbital periods are in years, while $\Delta U_{{\rm p}\rightarrow{\rm f}}$
and $r_{\rm frg}$ are in AU.  For example, a fragment separating at perihelion from
a sungrazer whose perihelion distance is 1.5~{\Rsun} and orbital period 750~yr ends
up in an orbit whose orbital period is almost exactly 900~yr when the fragment's
center of mass at breakup is 5~km farther from the Sun than the parent's center
of mass.  This effect is marginal or trivial for all comets other than sungrazers.

Another peculiar property of the Kreutz sungrazers is their anomalously low
orbital velocity at aphelion, of about 20~m~s$^{-1}$, very unusual at less
than 200~AU from the Sun.  As a result, a separation velocity of a few meters
per second could cause a fragment's orbital velocity to differ by more than
10~percent from the parent's, a sizable relative change that implies a major
orbit transformation.  And if fragmentation events should be distributed
randomly in time, their occurrence at large heliocentric distance is strongly
favored.

%
\begin{table*}[ht]
\vspace{-4.2cm}
\hspace{0.5cm}
\centerline{
\scalebox{1}{
\includegraphics{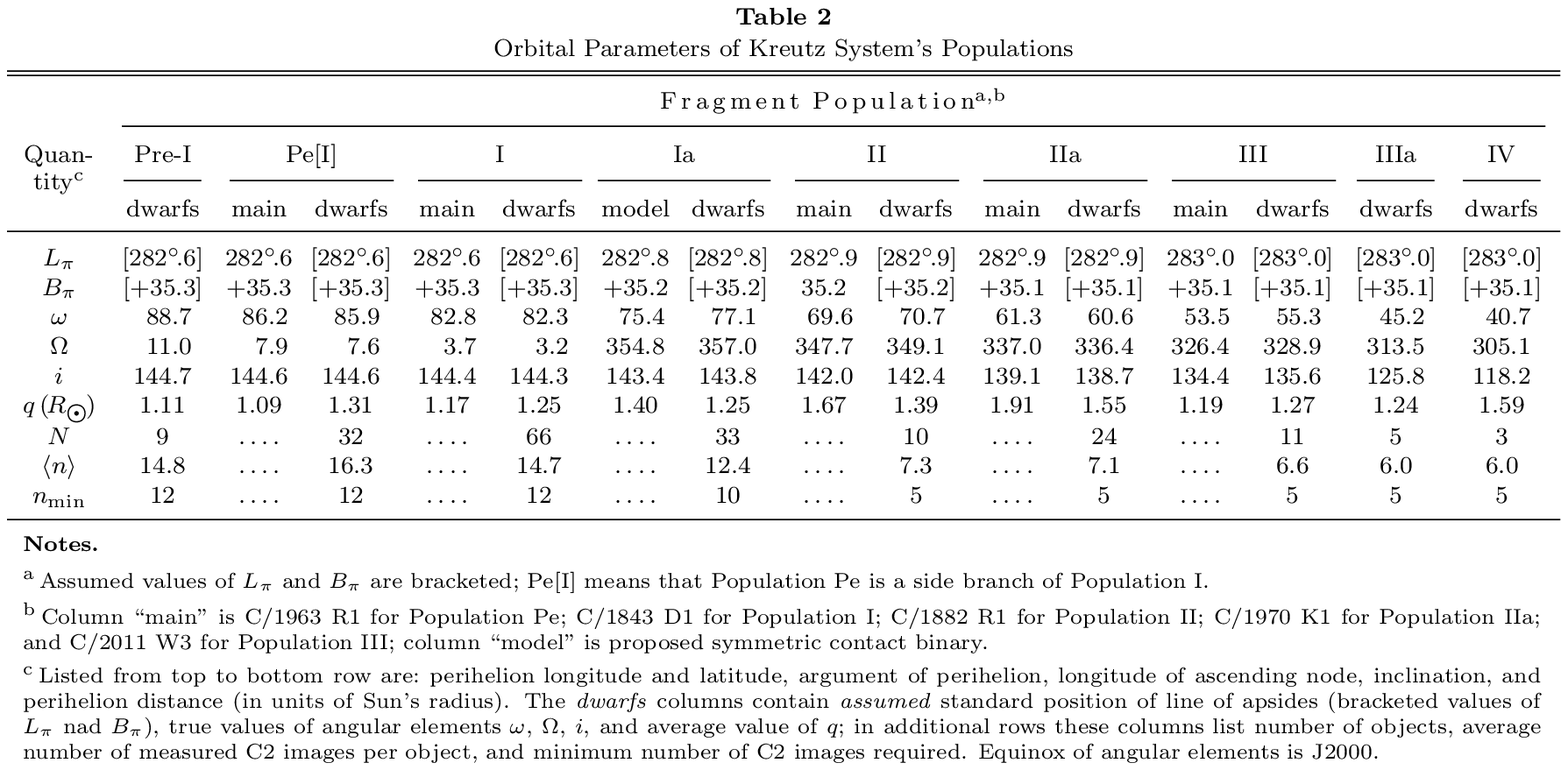}}} 
\vspace{-15.98cm}
\end{table*}

From the preceding it follows that fragments given birth in breakups
in close proximity of perihelion became members of the same population
but different clusters, whereas products of fragmentation events in proximity
of aphelion became members of the same clusters but different populations.  The
Great September Comet of 1882 and comet Ikeya-Seki (C/1965~S1) present an example
of two members of the same population that arrived at perihelion 83~yr apart,
while the sungrazers Pereyra (C/1963~R1), Ikeya-Seki, and White-Ortiz-Bolelli
(C/1970~K1) offer an example of three members of the same cluster that arrived
at perihelion within seven years of one another, but belonged to Populations~Pe,
II, and IIa, respectively.  

The bottom line is that the spread among the SOHO Kreutz sungrazers in the longitude
of the ascending node (a total of 66$^\circ$) and in the other true angular elements,
apparent from Table~2, can under {\it no circumstances\/} be explained by recurring
fragmentation at or near perihelion.  On the other hand, it was shown in Paper~1
--- and is briefly reviewed below --- that the contact-binary model offers a
self-consistent solution to the problem, fitting the observed data on the Kreutz
system.

\section{The Contact-Binary Model} 
The main features of the model are chronologically summarized in the following
subsections, which describe issues related, respectively, to (i)~the pre-split
progenitor; (ii)~the primary (initial) breakup into the two lobes (and the
neck); (iii)~secondary fragmentation of the lobes; (iv)~the arrival of the
first-generation fragments to perihelion; (v)~the arrival of the second- and
third-generation fragments to perihelion; (vi)~tidal fragmentation and dispersal
of fragments, including the SOHO dwarf sungrazers; and (vii)~the big picture
--- issues behind the major overhaul of the old hypotheses and the model
properties that withstood the challenges of new evidence.

\subsection{The Progenitor} 
The ultimate parent --- the progenitor --- of the Kreutz system in the recently
introduced model was Aristotle's comet.  The time of its appearance is usually
given as (ca.) 372~BC (e.g., Seargent 2009), but by combining all pieces of
information available and weeding out obvious errors, I believe that the
perihelion time of the comet --- if it indeed was a Kreutz sungrazer --- should
be known with accuracy better than $\pm$1~month.
%

It has to be remembered that the Greek year started with the month {\it
Hekatombaion\/} on the day of the first full moon after the summer solstice,
usually early July in our calendar.  After 683~BC it was customary to identify
the year by the name of the Athens' annual chief magistrate, called {\it
eponymous archon\/}.\footnote{See,{\vspace{-0.03cm}} e.g.,
{\tt https://www.hellenicaworld.com/Greece/History/}\\[0.05cm]
{\tt en/ArchonsOfAthens.html.}}
A parallely defined time scale was of course given by the Olympiad and its
year (from 1 to 4); for example, the year starting in July 373~BC and extending
through the end of June 372~BC was the 4th year of the 101st Olympiad, during
which Asteius was the archon at Athens.  Next came the 1st year of the 102nd
Olympiad, with Alcisthenes as archon.

The principal source of data on Aristotle's comet was his {\it Meteorologica\/}
1.6 (translated by E.\,W.\,Webster), where the object is mentioned twice, in
consecutive paragraphs.  In the first Aristotle says that ``the great comet,
which appeared at the time of the earthquake in Achaea and the tidal wave,
rose due west.''  In the next paragraph the text continues by repeating that
``the great comet we mentioned before appeared to the west in winter in frosty
weather \ldots, in the archonship of Asteius.''

%
\begin{figure}
\vspace{-0.65cm}
\hspace{2.85cm}
\centerline{
\scalebox{0.865}{
\includegraphics{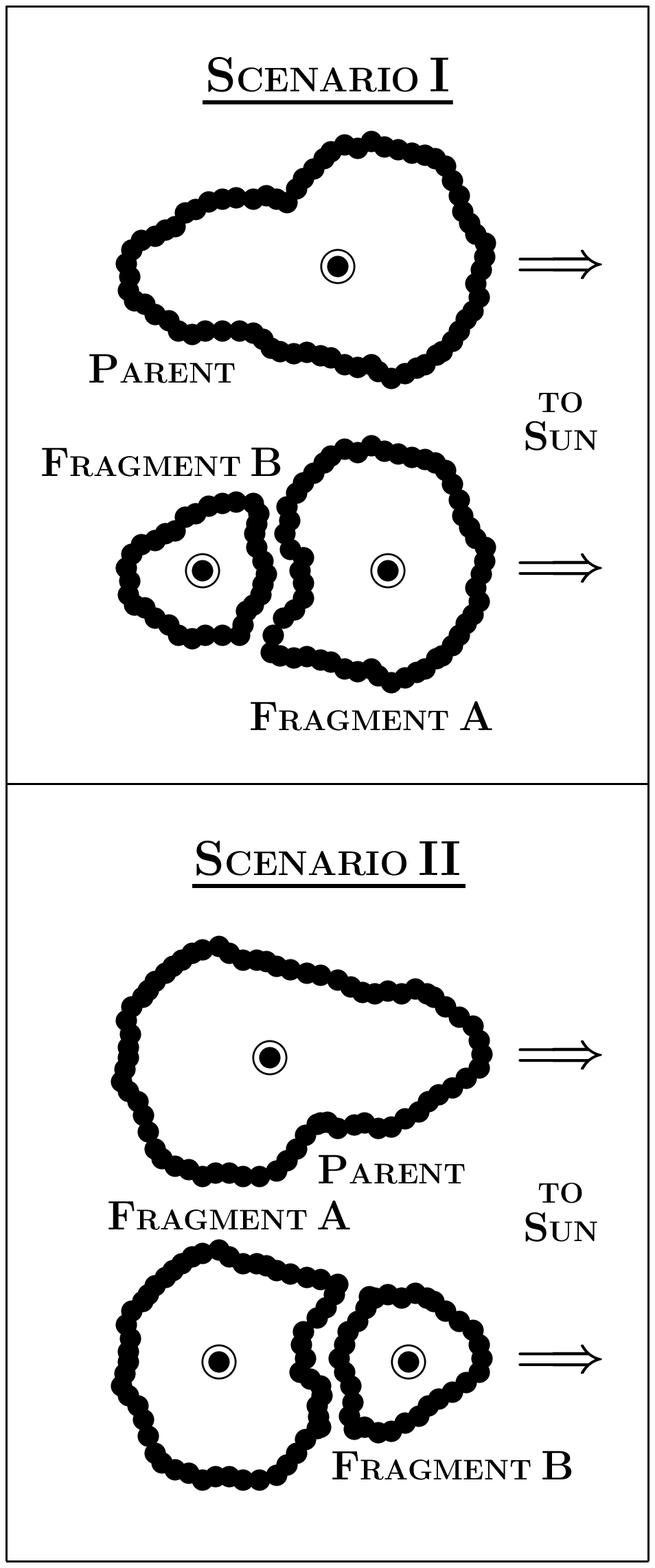}}} 
\vspace{-5.33cm}
\caption{Prolate cometary nucleus of a sungrazer shortly before and after
breaking up tidally into two uneven fragments at perihelion.  Turned to
the Sun (to the right) at the time of breakup is the larger end of the
nucleus, to become the primary fragment A, in Scenario~I, but the smaller
end, to become the escondary fragment~B, in Scenario~II.  The large
circled dots are the positions of the centers of mass of the parent
(or pre-split) nucleus, at the top of either panel, and of the fragments
A and B at the bottom.  The orientation relative to the Sun alone
assures that fragment A ends up in an orbit of a shorter period and
fragment~B in an orbit of a longer period than was the parent in
Scenario I, while the opposite is true in Scenario II. (Adapted from
Paper~2.)}
\end{figure}

The last sentence offers a tight constraint~on~the~time.  A Kreutz sungrazer is
seen (after sunset){\vspace{-0.03cm}} in~the~west~from Europe only when perihelion
occurs in February\,or\,later.\footnote{Comet C/1887~B1 (perihelion on 11~January)
was reported~only from the southern hemisphere, C/1880~C1 (perihelion on 28~January)
possibly also from China (Strom 2002); neither from Europe.}  On the other hand,
frosty weather in Greece seldom~extends into March (except in the
mountains).~For~a~Kreutz sungrazer, Aristotle's account alone shows that it was
at perihelion (and under observation) in February or (early) March.
Asteius' archonship means the year was 372~BC.

Before I address the time relationship of the Achaea earthquake, I compare
Aristotle's description with the story written by Diodorus Siculus about three
centuries later.  In his {\it Bibliotheca Historica\/} 15.50 (translated by
C.\,H.\,Oldfather) Diodorus says that ``[w]hen Alcisthenes was archon at
Athens \ldots and the Eleians celebrated the 102nd Olympiad \ldots, a divine
portent foretold the loss of [Lacedaemonians'] empire; for there was seen
in the heavens during the course of many nights a great blazing torch which
was named from its shape a `flaming beam,' and a little later \ldots the
Spartans were defeated in a great battle and irretrievably lost their
supremacy.''

The first part of this text is in glaring contradiction to Aristotle's account,
which obviously ought to be preferred.  Yet, one should ask why the conflict?
The answer appears to be furnished by the rest of Diodorus' narrative.  The
reference is to the Battle of Leuctra, which was fought on 6 July 371~BC and
which Spartans lost to Boeotians led by Thebans.  Moving the sighting of the
comet forward by one year, it was by that amount of time closer to the time
of the battle, preceding it by 4--5~months rather than 16--17~months.  That
of course made the comet a much more relevant portent.  How convenient!  This
sort of ``adjustment'' was unfortunately no exception among {\it some\/}
ancient historians.  In any case, Diodorus offered no good reason for
correcting Aristotle!

Now on the issue of the time that Aristotle assigned to the Achaean cataclysmic
earthquake that engulfed the city-states of Helike (submerged by a tsunami)
and Boura (collapsed and caved in).  The universally adopted time for the
earthquake has been the winter of 373~BC (e.g., Katsonopoulou 2017), which
depends in part on an equally universally adopted statement that the earthquake
occurred two years before the Battle of Leuctra.  Given that the battle was
in the summer and the earthquake in the winter, the statement makes no sense.
The two events must have been either 1$\frac{1}{2}$ or 2$\frac{1}{2}$~years
apart.  Since both Diodorus and Pausanias claimed that at the time of the
earthquake Asteius was the archon at Athens, it was the winter of 373/372~BC;
in the previous year the archon at Athens was Socratides.  This allows one to
state explicitly that the earthquake was apparently in December (or possibly
late November) 373~BC rather than January or February 373~BC.  If Aristotle's
statement that the comet appeared {\it at\/} the time of the earthquake was meant
to suggest that the two events occurred in the same winter, under the same archon
at Athens, then it was factually correct.  With Diodorus' time of the comet's
appearance discredited, the uncertainty of the perihelion time indeed
is clearly less than $\pm$1~month, and the comet was with high probability
observed during the months of {\it Anthesterion\/} and/or {\it Elaphebolion\/}
of the 4th year of the 101st Olympiad.

The contact-binary model is insensitive to the perihelion time of Aristotle's
comet, for which a standard date of $-$371.0 was adopted in Paper~2.  This implied
an orbital period of 735~yr, while the perihelion distance amounted to approximately
1.46~{\Rsun}, fairly close to the working model in Table~2, and the longitude of
the ascending node came out to be about 345$^\circ\!$.4, almost 10$^\circ$ less
than assumed for the working model.  The line of apsides has been holding remarkably
steady over the period of 24~centuries:\ in 372~BC the perihelion longitude was
282$^\circ\!$.38, the perihelion latitude +35$^\circ\!$.51.

\subsection{The Primary (Initial) Breakup} 
This is of course the point of ultimate importance --- the birth of the Kreutz system.
At issue is the origin of the sizable gap between the orbits of Populations~I and II,
or, more specifically, between the orbits of the Great March Comet of 1843 and the
Great September Comet of 1882 --- nearly 20$^\circ$ in the nodal longitude, 0.5~{\Rsun}
in the perihelion distance, etc.  Early on --- shortly after 1882 --- it was suspected
that the orbital motions of the sungrazers could have been affected differently by
drag in the solar corona (e.g., Elkin 1883), which was thought might explain the gap.
Besides, both the 1843 and 1880 sungrazers were measured only after perihelion.

The orbital gap was recognized as a serious problem once Kreutz (1891) found that the
preperihelion observations of the single nucleus and the post-perihelion observations
of the nuclear fragments of the 1882 sungrazer could readily be linked, ruling out a
perceptible effect of the solar corona.  However, little was done until Marsden's
(1967, 1989) thorough investigation of the orbital gap in terms of a long-term
influence of the indirect planetary perturbations.  Though ingenious, this hypothesis
required highly nonrandom perturbations acting over long periods of time, a scenario
that is contrary to evidence, as shown below.

%
%
\begin{figure}[t]
\vspace{-8.13cm}
\hspace{1.73cm}
\centerline{
\scalebox{0.76}{
\includegraphics{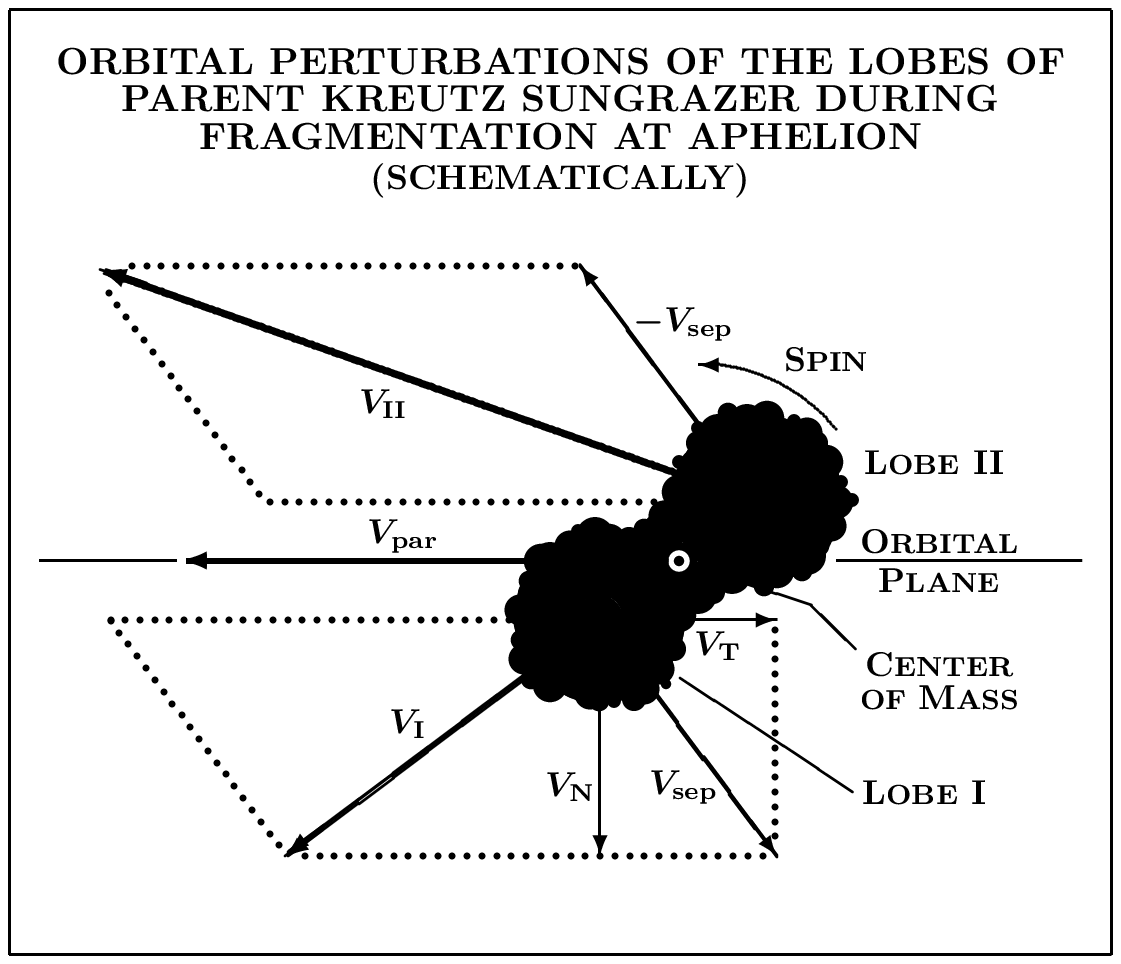}}} 
\vspace{-7.22cm}
\caption{Schematic representation, at the time of breakup, of the Kreutz system's
parent (progenitor) nucleus, modeled as a rotating contact binary consisting of
Lobe~I, Lobe~II, and the connecting neck.  The view is from the direction of the
Sun and the position of the original orbital plane is defined by the parent's
pre-breakup orbital velocity vector {\boldmath $V_{\bf par}$}.  The dot in the
middle of the neck is the center of mass of the parent, coinciding with the
projected spin axis, which at aphelion is assumed to point at the Sun.  As a result
of the breakup, the lobes are subjected to orbital perturbations.  At the time of
breakup the comet rotates counterclockwise, so that Lobe~I is released to the lower
right, moving relative to the center of mass in the direction of the separation
velocity vector {\boldmath $V_{\bf sep}$}, while Lobe~II is released to the upper
left, moving in the direction of the separation velocity vector {\boldmath $-V_{\bf
sep}$}.  The separation velocity consists of its transverse, {\boldmath $V_{\bf T}$},
and normal, {\boldmath $V_{\bf N}$}, components, the radial one is assumed to be
\mbox{\boldmath $V_{\bf R}$ = 0}.  Summed up with the parent's pre-breakup
orbital-velocity vector, the separation velocities insert Lobe~I into a new orbit
defined by the velocity vector {\boldmath $V_{\bf I}$} and Lobe~II into an orbit
defined by the velocity vector {\boldmath $V_{\bf II}$}.  The Great March Comet
of 1843 is the largest surviving mass of Lobe~I, the Great September Comet of
1882 is the largest surviving mass of Lobe~II.  For clarity, the orbital and
separation velocities are not drawn to scale; in the scenario in Paper~1 the
ratio \mbox{{\boldmath $|V_{\bf sep}|$}/{\boldmath $|V_{\bf par}|$} = 0.13}.
(Reproduced from Paper~2.){\vspace{0.45cm}}}
\end{figure}

Unlike a gradual buildup of minor contributions, the scenario proposed to
explain the orbital gap between the 1843 and 1882 sungrazers in the context of
the contact-binary model in Paper~1 was by a brief event in which the progenitor
split into the two lobes it consisted of, with the connecting neck possibly
as a third fragment.  To simplify the matter, a scenario of overall symmetry,
with the spherical lobes of equal size and density, was adopted.  The
modeled event was assumed to have occurred at aphelion on 30~December~5~BC
and at about 163~AU from the Sun, which implies the Kreutz system's age of
almost exactly two millennia.  A schematic representation at the instant
of the breakup is in Figure~5; in the view from the Sun the progenitor
rotated counterclockwise along the axis directed toward the Sun.  The
rotation velocity measured at each lobe's center of mass was the separation
velocity, which, added vectorially with the progenitor's orbital velocity,
determined the lobe's orbital velocity and upcoming orbit.  The three
angular elements were affected by the normal component of the separation
velocity, the perihelion distance by the transverse component, which also
had a marginal influence on the next perihelion time in the sense that the
fragment subjected to a transverse separation velocity in the direction
of the orbital velocity should arrive at perihelion days later.

In the initial breakup, Lobe I, the protofragment of Population~I, separated
from Lobe~II, the protofragment of Population~II, in line with the stipulations
that the Great March Comet of 1843 was the largest surviving mass of Lobe~I
and the Great September Comet of 1882 the largest surviving mass of Lobe~II.

The working model, updated in Sekanina (2022; hereafter referred to as Paper~3),
shows the relevant separation velocity{\vspace{-0.02cm}} of Lobe~I had a normal
component of $-$1.80~m~s$^{-1}$ and a transverse component of $-$1.86~m~s$^{-1}$,
the opposite for Lobe~II.  The respective corrections to the progenitor's orbit
for Lobes~I and II were +7$^\circ\!$.3 and $-$5$^\circ\!$.7 in the argument
of perihelion; +9$^\circ\!$.0 and $-$7$^\circ\!$.1 in the longitude of the
ascending node; +1$^\circ\!$.0 and $-$1$^\circ\!$.4 in the inclination;
$-$0.23~{\Rsun} and +0.27~{\Rsun} in the perihelion distance; and $-$1.31~day
and +1.56~day in the time of the next perihelion passage (Section 4.4).
Accordingly, the difference Lobe~II minus Lobe~I was predicted to have amounted
to $-$13$^\circ\!$.0 compared to $-$13$^\circ\!$.2 in Table~2 in the argument
of perihelion; and $-$16$^\circ\!$.1 compared to $-$16$^\circ\!$.0 in the
longitude of the ascending node.  The differences of $-$2$^\circ\!$.4 in the
inclination and +0.50~{\Rsun} in the perihelion distance have been fitted
perfectly.

\begin{figure*}[t]
\vspace{-11.58cm}
\hspace{-0.2cm}
\centerline{
\scalebox{0.84}{
\includegraphics{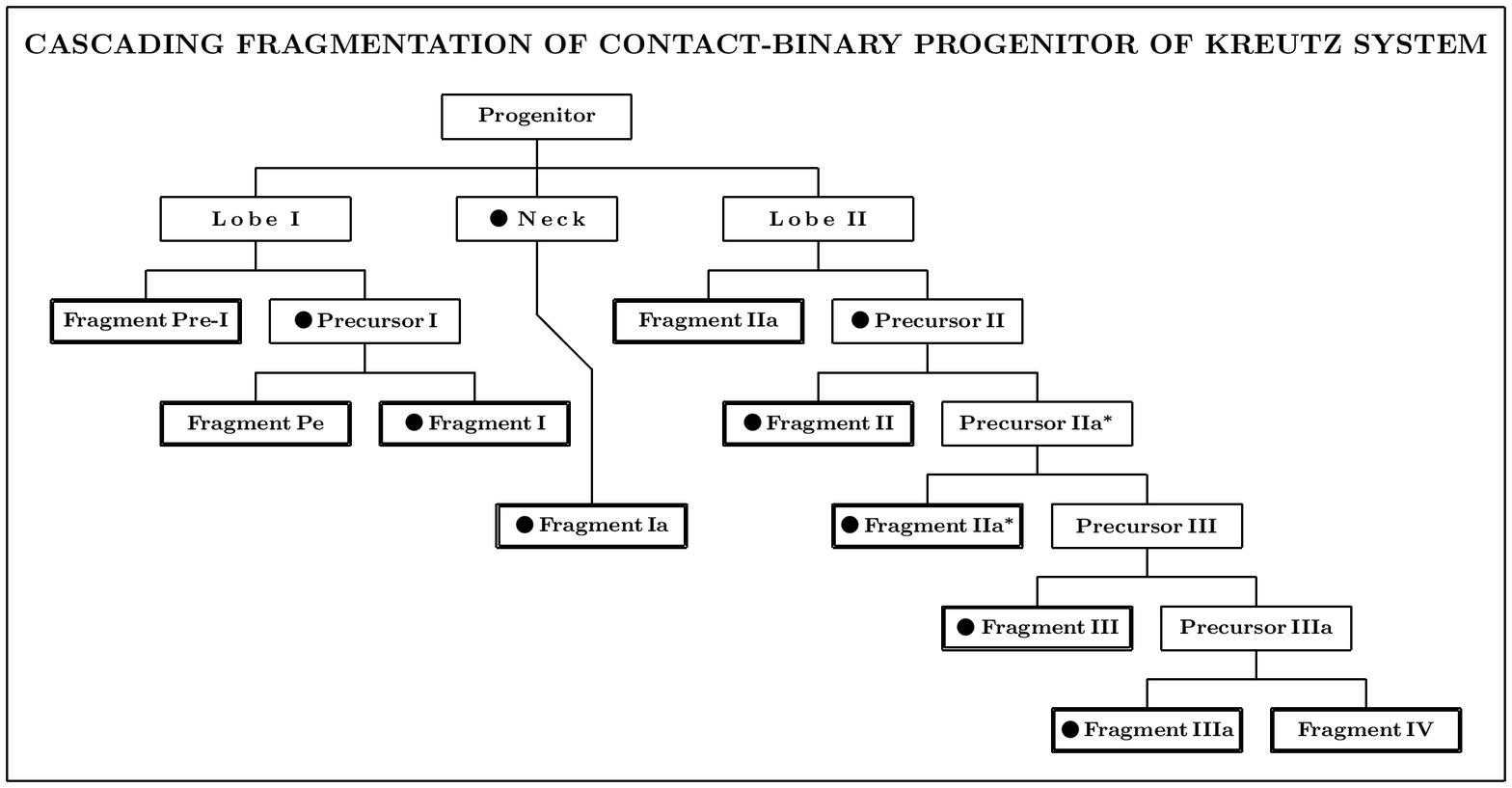}}} 
\vspace{-4.27cm}
\caption{Pedigree chart for cascading fragmentation of the contact-binary
Kreutz progenitor.  The bullets show the fragments that the working model
in Paper~1 assumed to move in the same orbits as their immediate parents.
The sungrazers in the heavily framed boxes reached their first perihelion as
the daylight comets of AD~363 (Section~4.4).  Their orbits are approximated
as follows:\ Fragment~I by C/1843~D1; Fragment~II by C/1882~R1; Fragment~Pe
by C/1963~R1; Fragment~IIa by C/1970~K1; and Fragment~III by C/2011~W3.  The
orbits for Fragments~Ia, Pre-I, IIa$^\ast\!$ (see Section~5), IIIa, and IV are
described elsewhere in this paper. (Reproduced from Paper~3.){\vspace{0.9cm}}}
\end{figure*}

Similarly well was fitted the simulated orbital solution to the progenitor's
fragmentation, derived by numerically integrating the orbits of the 1843
and 1882 sungrazers in Paper~2.  In this solution as well as in the woking
model, the neck connecting the two lobes was assumed to have acquired
no extra velocity at the breakup, continuing to move in the orbit of the
progenitor before splitting.  This motion is characterized in Table~2 by
the orbital elements of Population~Ia in the column {\it model\/}.  Since
there is no major, naked-eye sungrazer known to move in the orbit of
Population~Ia and since the deviations between the {\it model\/} and
{\it dwarfs\/} sets of elements in Table~2 are fairly sizable, it is
possible that the neck remained attached to one of the two lobes after
the initial breakup (probably Lobe~I) and separated from it later.

The radial component of the separation velocity was assumed to be zero in
both the working model (Papers~1 and 3) and the simulated orbital solution
(Paper~2).  This component affects the next perihelion time much more
significantly than the transverse component; for a fragmentation event at
aphelion an increase in the radial velocity by 0.1~m~s$^{-1}$ would move
the next perihelion time by 5.2~days.

The final point in this subsection concerns the dependence of fragments'
orbits on the distance of the fragmentation event from aphelion.  The
choice of aphelion was conditioned on the fact that the separation velocity
needed is always a minimum.  The rate of increase in the velocity needed
to achieve the same effect 130~years before or after aphelion (150~AU from
the Sun) is 10~percent in the argument of perihelion and the longitude of
the ascending node, 5~percent in the inclination, and 8~percent in the
perihelion distance.  The rate varies at an accelerated rate, the numbers
at 230~years before or after aphelion (120~AU from the Sun) being,
respectively, 40~percent, 23~percent, and 32~percent.  At a zero radial
separation velocity, the change in the next perihelion time increases
by about 2~days when the fragmentation event is moved from aphelion to
230~years before perihelion, but decreases by about 1~day when it is moved
to 230~years after perihelion.  In general, the fragmentation solutions
are found to vary insignificantly over a time scale of a century or so
from the aphelion time. 

\subsection{Secondary Fragmentation of the Lobes (and Neck)} 
The dwarf Kreutz sungrazers in Populations I, II, and possibly Ia, are
explained as fragmentation end products of the major surviving masses of,
respectively, Lobes~I, II, and perhaps the neck.  But what about the
dwarf~sun\-grazers in the other{\vspace{0.03cm}} populations?

Given the cascading nature of the fragmentation process, the chance was
that the separated lobes continued to split.  And because of the nature
of fragmentation far from the Sun, one can reasonably expect that the
secondary events took place under essentially the same conditions as
the primary breakup, except for decreasing dimensions{\vspace{0.03cm} of the objects.

The sequence of proposed fragmentation events is presented in a pedigree
chart in Figure~6, copied from Paper~3.  They are expected to have occurred
--- in the depicted order from the top down --- in the aphelion region of
the progenitor's orbit.  The exception was Fragment Pe, a precursor of
comet Pereyra, which apparently separated from Lobe~I later, at about 70~AU
on the way to perihelion.  In line with the proposed classification of
the Kreutz system, it is contemplated  that, repeatedly, a parent fragment
split into two fragments, one of which containing most mass became a new
parent fragment only to split again.  The fragments in the heavy boxes in
Figure~6 were on the way to their first perihelion and are in the following
referred to as{\vspace{0.03cm} the {\it first-generation fragments\/}.

Figure~6 shows that fragmentation proceeded in the direction of decreasing
true nodal longitude for both lobes.  At the end of the Lobe~II sequence,
Precursor~III split into Fragment~III (whose existence is supported by comet
Lovejoy C/2011~W3 and by Population~III dwarf comets) and Precursor~IIIa,
which in the final documented fragmentation step split into Fragments~IIIa
and IV.  The total separation velocities of the five fragments from IIa
to IV were, respectively, 3.5~m~s$^{-1}$, 3.3~m~s$^{-1}$, 3.2~m~s$^{-1}$,
4.4~m~s$^{-1}$, and 4.6~m~s$^{-1}$ (Paper~3).  Lobe I fragmented less profusely.
Its main product was Fragment~I, others were Fragment~Pre-I, whose derived
separation velocity was 1.5~m~s$^{-1}$ and, centuries later, Fragment~Pe,
separating with a velocity of 2.4~m~s$^{-1}$.  Given that the separation
velocities were essentially constant, the parent fragments were progressively
spun up, as their dimensions were getting smaller.

\subsection{First-Generation Fragments at Perihelion\\in AD~363} 
One trait of Aristotle's comet that made it an attractive candidate for
the Kreutz progenitor was its timing:\ the period of time between its
appearance and the appearance of the Great Comet of 1106 (X/1106~C1),
a likely Kreutz sungrazer, was almost exactly twice the presumed orbital
period of the Great March Comet of 1843, whose previous return to perihelion
the 1106~comet was believed to portray.  The weak point of this argument
had been the absence of a related event halfway between 372~BC and AD~1106,
most probably in the decade of AD~360--370.  Already in the aftermath of
the 1882 sungrazer's appearance, this problem was noticed by Hall (1883),
who proposed a link between Aristotle's comet, a {\it missing\/} comet in
AD~368, and the comets of 1106 and 1843.  Similarly, the two superfragments
were predicted by Sekanina \& Chodas (2004) to pass perihelion one week apart
in AD~356.  The perihelion return of two or more sungrazers around this time
has become an indispensable condition, following a postulated near-aphelion
fragmentation event.

Because of its confused presentation in the literature, I had long dismissed
a note written in the treatise {\it Res Gestae\/} by Ammianus Marcellinus,
a prominent Roman historian of the 4th century AD, that in the year 363
{\small \bf ``in broad daylight comets were seen''} (Rolfe 1940) from Antioch
on the Orontes, the capital of the Syrian Province of the Roman Empire.  For
example, Ramsey (2007) refers to this event as the ``Jovian comet'' and
dates it August--September 363.  Ho (1962), whose catalogue comprises only
Far Eastern sources, says (under \#173) that a {\it po\/} (tailless) comet
was seen between August~26 and September~23 first in Virgo (near $\alpha$~Vir
and $\zeta$~Vir) and then it entered Hercules, Aquila, etc., so it could
under no circumstances have been a sungrazer; also, in August--September it
was not a daytime comet.  Only after reading about Ammianus' account in the
carefully researched book by Seargent (2009), in which he actually suggested
that this may have been a case of Kreutz sungrazers, it became clear to me
that the Chinese comet in Ho's catalogue had nothing in common with the event
commented on by Ammianus.  And although the historian offered no dates, his
narrative suggests that the observation was most probably made in the course
of November~363.  Most significantly, the text specifically says {\it comets\/}
rather than {\it a comet\/} and the {\it daylight\/} sighting implies that
the comets were extraordinarily luminous, probably around apparent magnitude
$-$10 or brighter and located near the Sun both in the sky and in space.
This must have been an unprecedented, spectacular celestial show that, as
far as I know, was never to be witnessed again!

The brief comment by Ammianus provokes a number of questions.  Why the
sighting was not reported by anyone else?  Why not by the Chinese?  Can
one be sure that the timing of the sighting of the comets as a portent
was not manipulated in relation to historical events?  What has been Ammianus'
professional reputation?  How many comets were seen?  And over what
period of time?

Some of these questions can be answered more or less fully, some partially,
and some not at all.  First, the work of Ammianus has generally been praised
(e.g., Matthews 2008); he almost certainly was the top Roman historian of
his time, was knowledgeable in natural sciences, and his description of the
earthquake and tsunami of AD~365 in Alexandria has been deemed factually
accurate.

On the other hand, Ammianus was a protagonist of paganism and believed in omens
and portents.  Because of these tendencies, he greatly preferred emperor Julian to
Jovian, even though he served under both.  If Ammianus should have manipulated the
date of sighting of the comets (as Diodorus had in the case of the comet of 372~BC),
he would have moved the event to early 363 to present it as a portent of the
forthcoming death, in June 363, of his beloved pagan emperor Julian, rather than
the pro-Christian Jovian.  He did not.

If the Chinese had observed the daytime event in AD~363, they would have
recorded it as {\it sun-comets\/} to use Strom's (2002) terminology.  At first
sight, the absence of a Chinese account is surprising.  However, Strom, who
examined the statistics of the sun-comets in the Chinese annals, discovered that
the sightings were strongly concentrated in the summer months in spite of summer
monsoon.  Besides, there was only one sun-comet record before 1539.  The lack
of corroboration of Ammianus' account from China may not be too
unexpected~after~all.\,\,\,\,\,\,

It is unfortunate that Ammianus failed to provide more information on the daylight
comets.  One can only speculate about their number and the period of time involved.
Based on the working model in Paper~3, which consisted of 10~sizable fragments whose
perihelion times in AD~363 spanned 4.4~days, one can guess that up to 70~percent of
the fragments could have been above the naked-eye detection threshold at the most
favorable time, as illustrated in Figure~7.

Once the first-generation fragments reached perihelion in AD~363, the tidal
fragmentation began to multiply the numbers and thereby to make the structure
of the Kreutz system much more complex.  To offer the potential evolutionary
path for any individual sungrazer requires the knowledge of a rather accurate
set of its orbital elements, including the period.  Unfortunately, this
information is available for only a few Kreutz comets.

\begin{figure*}[t]
\vspace{-2.25cm}
\hspace{-0.05cm}
\centerline{
\scalebox{0.865}{
\includegraphics{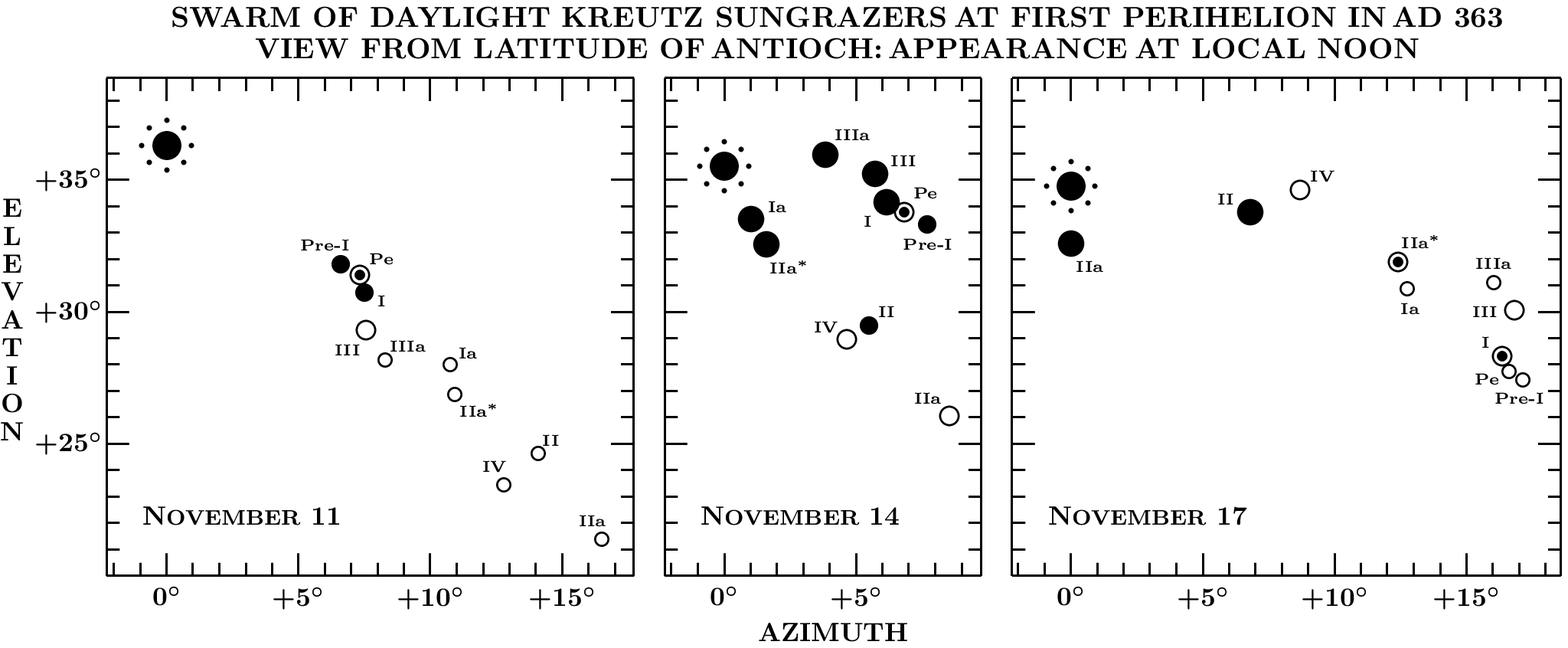}}} 
\vspace{-16.35cm}
\caption{Simulation of the swarm of ten Kreutz sungrazers in projection onto the
plane of the sky, viewed, near the Sun, from the latitude of Antioch at local
noon on 363 November 11, 14, and 17.  The Sun is plotted near the upper left
corner.  The horizontal coordinate system has the azimuth reckoned positive
from the south through the west, the elevation reckoned positive from the
horizon to the zenith.  The symbols depict each sungrazer's visibility to
the naked eye, defined by a visibility index $\Im$ (in magnitudes):\ large
solid circles stand for \mbox{$\Im > +1.5$} (clearly visible); medium solid
circles for \mbox{$+0.5 < \Im \leq \! +1.5$} (probably visible); circled dots
for \mbox{$-0.5 \leq \Im \leq \! +0.5$} (potentially visible); medium open
circles for \mbox{$-1.5 \leq \Im < \! -0.5$} (probably invisible); small open
circles for \mbox{$\Im < \! -1.5$} (invisible); and small diamonds for fragments
that are below the horizon.  See Paper~3 for additional information. (Reproduced
from Paper~3.){\vspace{0.95cm}}}
\end{figure*}

\subsection{Second- and Third-Generation Fragments} 
In the continuing story of the computer simulated evolution of the Kreutz system
in Paper~2, Fragment~I --- the largest fragment of Lobe~I --- broke at the 363
perihelion into at least two subfragments.  The largest piece, whose orbital
motion was closely simulated, ended up in a slightly changed orbit with a period
of 742.2~yr and returned to perihelion as the Great Comet of 1106.  With very
minor orbital changes at this perihelion it then completed another revolution
about the Sun to return as the Great March Comet of 1843. 

Similarly, Fragment~II --- the largest fragment of Lobe~II --- appears to have
split tidally at perihelion in 363 into a number of subfragments.  The largest
subfragment entered a weakly modified orbit with a period of 774.7~yr and returned
to perihelion as a Chinese comet of 1138, \#403 in Ho's (1962) catalogue.  As
modeled in Paper~2, the comet suffered at least two breakups in 1138, the last one
--- presumably hours after perihelion --- producing the famous pair of the Great
September Comet of 1882 and Ikeya-Seki of 1965.  A comprehensive orbital reanalysis
of component~A of Ikeya-Seki in Paper~2 was instrumental in establishing beyond
doubt that this comet previously appeared in the late 1130s, decades after the
arrival of the Great Comet of 1106.  This result is in excellent agreement with
Marsden's (1967) computation of the previous appearance of the brightest nucleus
of the 1882 sungrazer and in line with his conclusion that the two comets separated
from a shared parent at perihelion in the 12th century.

Having split off from Precursor~I on the way to the 363 perihelion (Figure~6),
Fragment~Pe was presumed to have likewise undergone an event of tidal fragmentation
in AD~363, one of its subfragments having returned to perihelion in early August
1041.  According to Hasegawa \& Nakano (2001), this comet --- a suspected Kreutz
sungrazer --- was observed in China and Korea from the beginning of September
1041 on, its celestial path resembling that of the Chinese comet of 1138.  Comet
Pereyra of 1963 was proposed in Paper~1 as one of the fragments of the 1041
comet.

The second-generation fragments --- the comets of 1041, 1106, and 1138 --- and
the third-generation fragments --- the bright 1843 and 1882 sungrazers, Pereyra,
and Ikeya-Seki --- complete the set of members of the Kreutz system, whose
orbital evolution was successfully simulated in Paper~2 starting with
the progenitor two millennia ago.  Although the orbital period is also known
with high accuracy for comet Lovejoy, C/2011~W3 (Sekanina \& Chodas 2012), a
sungrazer associated with Population~III, it was not possible to link this
object with any obvious candidate of the Kreutz system in the past.  For other
naked-eye sungrazers of the 19th and 20th centuries, the orbital period was not
determined:\ White-Ortiz-Bolelli of 1970 (associated with Population~IIa), the
Great Southern Comet of 1880 (C/1880~C1), the Great Southern Comet of 1887
(C/1887~B1), and of course the eclipse comet of 1882 (X/1882~K1) --- all
associated with Population~I or Pe.  Populations Pre-I, Ia, IIIa, and IV are
related to no known naked-eye Kreutz sungrazer.

The sets of orbital elements for the progenitor, the lobes, and the fragments
of the three generations, as derived in Paper~2, are summarized here in Table~3.
Listed for the progenitor are both the heliocentric elements at perihelion in
372~BC and the barycentric elements at the time of the aphelion breakup in 5~BC.

\begin{table*}[t]
\vspace{-4.15cm}
\hspace{0.55cm}
\centerline{
\scalebox{0.99}{
\includegraphics{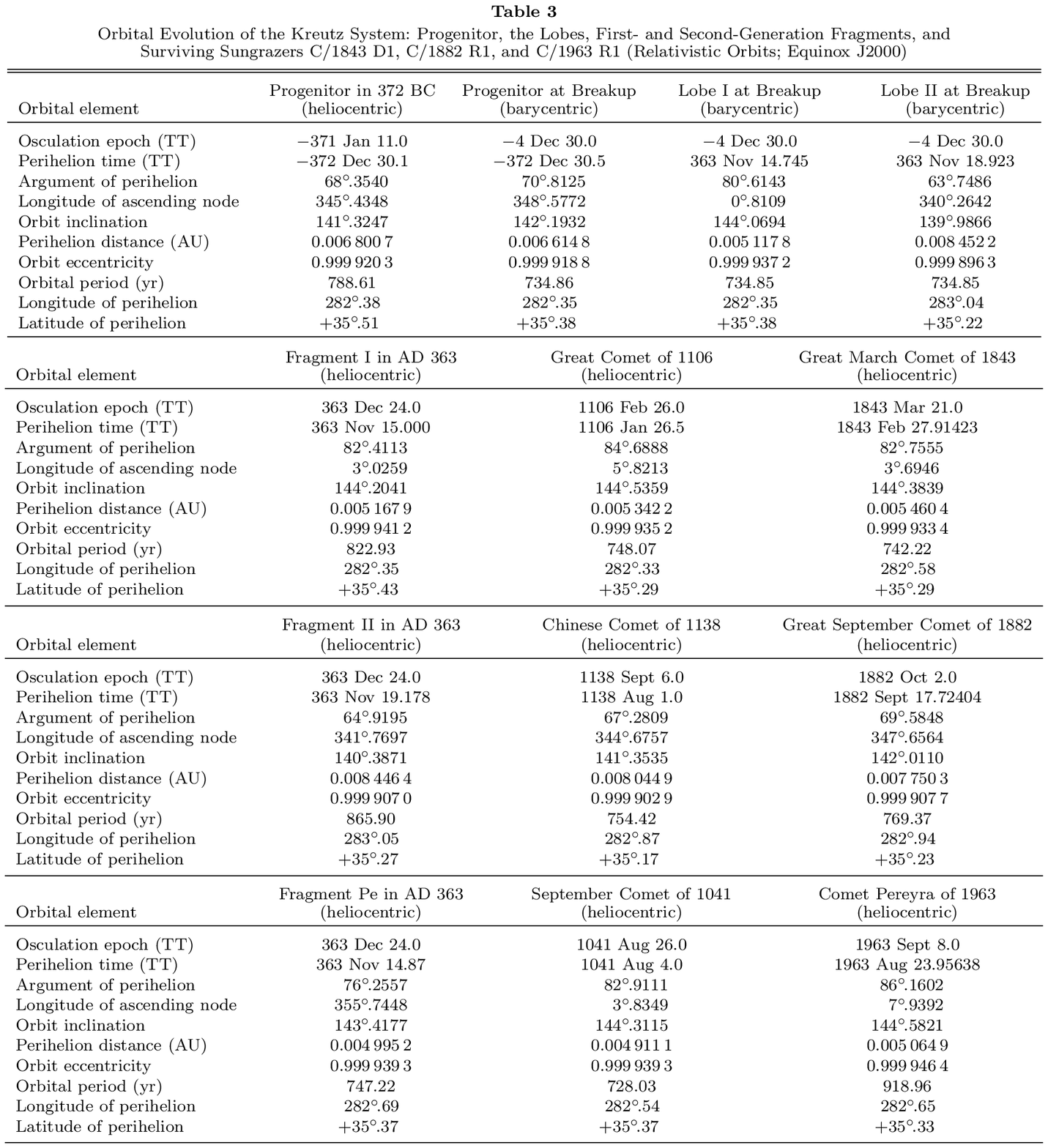}}} 
\vspace{-5.38cm}
\end{table*}

\subsection{Tidal Fragmentation and Random Dispersal\\of Fragments}
%
The events of tidal fragmentation at perihelion are extremely powerful in their
ability to randomly disperse the fragments in time and space.  I present
a few examples to illustrate the magnitude of these effects on the orbital
period, governed by a shift of the center of mass along the radius vector,
$\Delta U_{{\rm p} \rightarrow {\rm f}}(r_{\rm frg}, P_{\rm par})$, expressed
by Equation~(4) in Section~3.

The first case is the evolution of the well-behaving Great March Comet of 1843
and its Population~I precursors.  Fragment~I did apparently break up into a
number of subfragments at the 363~perihelion, the most sizable of which became
the Great Comet of 1106, whose center of mass shifted merely by \mbox{$\Delta
U_{{\rm p} \rightarrow {\rm f}}(q) = -0.70$ km}.  This indicates that should
Fragment~I have not split, it would have reached perihelion in AD~1139.  This
negative shift of the 1106 comet must have been balanced by positive shifts of
greater magnitudes for smaller naked-eye fragments due to arrive at perihelion
at least a century later.  The comet of 1232, a Kreutz candidate according to
Hasegawa \& Nakano (2001), is a case in point.

Much more complicated was the evolution of the Great September Comet of 1882 and
its precursors.  Fragment~II must have broken into a larger number of subfragments
than Fragment~I.  The center-of-mass shift of the Chinese comet of 1138, the
putative parent to the 1882 sungrazer and Ikeya-Seki, was formally computed to
equal $-$5.25~km, which implies that should Fragment~II have not split in AD~363,
it would have passed perihelion in AD~1237.  However, the computed shift value
is affected by the separation of Ikeya-Seki and the 1882 comet; the correct value
should be closer to zero and compensated by larger positive shifts of smaller
naked-eye sungrazers in the 14th and possibly even the 15th centuries; one of
Hasegawa \& Nakano's (2001) Kreutz suspects, the comet of 1381, is the example
of a plausible candidate.  And, significantly, some of the Population~II objects
should be due for another return to perihelion by the mid-21st century. 

The center of mass of a fragment of the comet of 1041 needed \mbox{$\Delta U_{{\rm
p} \rightarrow {\rm f}} = +3.78$ km} to fit the motion of comet Pereyra, as
otherwise it would have returned to perihelion in AD~1744.  One or more bright
siblings of comet Pereyra, perhaps of comparable size and prominence, may have
arrived in the 17th century, such as, say, C/1668~E1 and/or C/1695~U1.

The ultimate case of this kind of orbit dispersal has been documented by
the statistics of the dwarf sungrazers' motions, especially those imaged by
the SOHO coronagraphs.  Since the overwhelming majority of the SOHO dwarf
Kreutz comets belongs to Population~I, I focus in the following on this
particular set.

Because the SOHO dwarf sungrazers are known to never survive perihelion, their
age cannot exceed one revolution about the Sun and the time of their birth could
not predate the perihelion time of the Population~I parent, the Great Comet
of 1106.  I assume that these 10-meter-or-so sized fragments were released at
perihelion $q$ (in AU) from all over the comet's nucleus of diameter $D$ with
zero velocity, undergoing the same dispersion process as the major fragments.
At a bulk density of 0.5~g~cm$^{-3}$ the mass of a fragment is about
$3 \times 10^8$\,grams.  I further assume that the rate of release, measured
by the number of dwarf sungrazers returning to perihelion per year, varies
as the surface area of the parent nucleus.  For a perihelion breakup,
Equation~(4) shows that an infinitesimal increase of $dP_{\rm frg}$ in a
fragment's orbital period $P_{\rm frg}$ (in yr) is invoked by a differential
center-of-mass shift (in AU) along the radius vector of
\begin{equation}
du = {\textstyle \frac{1}{3}} q^2 P_{\rm frg}^{-\frac{5}{3}} dP_{\rm frg}. 
\end{equation}
For \mbox{$dP_{\rm frg} = 1$ yr}, Equation~(5) offers a range of radial shifts
$du$ of all fragments with the period near $P_{\rm frg}$ that arrive at
perihelion per year.  These fragments come from a ring area of the surface
whose center-of-mass shift along the radius vector from the nuclear center of
the parent comet, $\Delta U_{{\rm p}\rightarrow{\rm f}}$, is related to the
comet's orbital period, $P_{\rm par}$, by
\begin{equation}
\Delta U_{{\rm p}\rightarrow{\rm f}} = {\textstyle \frac{1}{2}} q^2 \!\!\left(
 \!P_{\rm par}^{-\frac{2}{3}} \!-\! P_{\rm frg}^{-\frac{2}{3}} \!\right) \!. 
\end{equation}
The size of an infinitesimal surface area is
\begin{equation}
dA = {\textstyle \frac{1}{2}} \pi D^2 \! \cos \theta \, d\theta, 
\end{equation}
where $\theta$ is the angle, measured from the nuclear center, that the
line connecting it with any point of the ring made with the plane normal
to the radius vector and \mbox{$du = \frac{1}{2} D \cos \theta \, d\theta$}.
Equation~(7) now becomes
\begin{equation}
dA = \pi D \, du = {\textstyle \frac{1}{3}} \pi q^2 D P_{\rm frg}^{-\frac{5}{3}}
 dP_{\rm frg}. 
\end{equation}

Now about the recent influx of the SOHO sungrazers:\ Let the annual rate of
their arrivals at a reference time be $\Delta {\cal N}$, coming from an area
$\Delta A$, and let their orbital period be $P_{\rm frg} = {\cal P}$.  Let the
number of SOHO-like sungrazers released from the whole surface of the comet's
nucleus in 1106 be ${\cal N}$.  Since the number is assumed to be proportional
to the surface area, I have at the reference time with help of Equation~(8)
\begin{equation}
\Delta {\cal N} = \zeta \Delta A = {\textstyle \frac{1}{3}} \zeta \pi
 q^2 D {\cal P}^{-\frac{5}{3}}, 
\end{equation}
where $\zeta$ is a constant of proportionality.  The total population has to
equal
\begin{equation}
{\cal N} = \zeta \pi D^2 
\end{equation}
and, eliminating $\zeta$,
\begin{equation}
{\cal N} = \frac{3D}{q^2} {\cal P}^{\frac{5}{3}} \Delta {\cal N}. 
\end{equation}
Here $D$ and $q$ are in AU, ${\cal P}$ in yr.  If the nucleus of the 1106 comet
was, say, 50~km across, the shift $\Delta U_{{\rm p} \rightarrow {\rm f}}$
relative to the center of mass varied from $-$25~km to +25~km.  The comet's
orbital period $P_{\rm par}$ equaled 737~yr (= 1843--1106) and the perihelion
distance amounted to 0.005342~AU (Table~3).  The shortest orbital period of
the fragments, corresponding to a shift of $-$25~km was 270~yr, the longest,
corresponding to +25~km, 79,000~yr.  The stream of the SOHO sungrazers is thus
predicted to have begun in the late 14th century and will go on for another
nearly 80~thousand years.  The last objects of the stream will reach nearly
3700~AU from the Sun at aphelion.

Let us take the year 2010 as the reference time.  The rate of arriving sungrazers
and sunskirters in the decade around that year equaled about 200~yr$^{-1}$, the
Kreutz comets made about 85~percent of the total, and Population~I alone contributed,
judging from Table~1, a little more than 50~percent of all Kreutz members.  I
take the rate of Population~I members to equal \mbox{$\Delta {\cal N} \simeq 100$
yr$^{-1}$}.  Their orbital period ${\cal P}$ averaged \mbox{2010--1106 = 904 yr},
so that the ratio \mbox{${\cal N}/\Delta {\cal N} \simeq 3000$} and the total
number of the Kreutz SOHO sungrazers in Population~I comes out to be approximately
300,000.  Their total mass is less than 10$^{14}$\,grams over the
nearly 80~thousand years, the mass of a subkilometer-sized cometary nucleus,
a trivial fraction of the expected mass of the 1106~comet.

\begin{table*}[t]
\vspace{-4.1cm}
\hspace{0.29cm}
\centerline{
\scalebox{0.97}{
\includegraphics{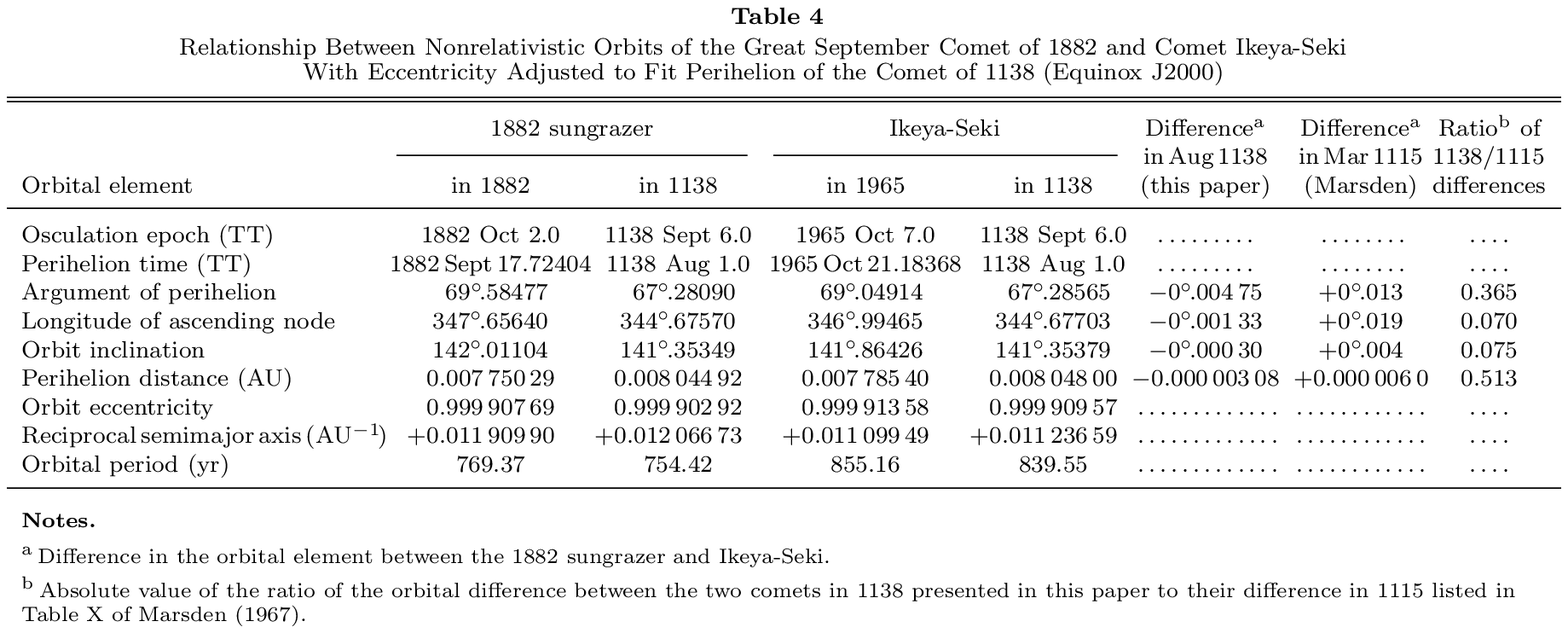}}} 
\vspace{-16.7cm}
\end{table*}

\subsection{The Big Picture:\ Issues Behind the Major Overhaul and Features Shared}
%
In Section 1 I already commented on some of the important developments in the
past 15~years that made the introduction of a new model imperative.  However,
the contact-binary model is not merely an update of the two-superfragment model
(Sekanina \& Chodas 2004), it works on a conceptually higher level.  While the
two-superfragment model was by and large {\it ad hoc\/} in its approach, the
contact-binary model has features of a {\it pyramidal\/} construct.  The 2004
model focused on the relationships among eight {\it bright, naked-eye sungrazers
only\/}, optimizing the locations of fragmentation events in order that the
separation velocities stay as low as possible.  This effort was not entirely
successful, because the values near \mbox{6--7}~m~s$^{-1}$ were common and the
initial breakup of the progenitor into the two superfragments required that they
detached at $>$8~m~s$^{-1}$.  The model also used the incorrect premise that
the Great Comet of 1106 was a member of Population~II, whose fragments included
both comet Pereyra and White-Ortiz-Bolelli, thus eliminating Populations~Pe and
IIa.  Aristotle's comet did not fit the progenitor and no serious effort was
made to incorporate the SOHO sungrazers; they were added only in part~2 of the
investigation (Sekanina \& Chodas 2007).

On the other hand, the contact-binary model skillfully integrates the data on
the naked-eye and SOHO dwarf members of the Kreutz system.  The greatly expanded
number of populations, whose existence is established by evidence from the SOHO
sungrazers, is shown to be the result of a stochastic process of cascading
fragmentation rather than ad hoc breakup events and the separation velocities
are kept below 5~m~s$^{-1}$.  The progenitor is equated with Aristotle's comet
and the SOHO sungrazers are logically presented as the end products of the
same process.  The orbital correlations between the naked-eye and coronagraphic
sungrazers are documented in the five cases when the former is known.  The model
also profits from the refined orbit determination of the 1843 sungrazer (Sekanina
\& Chodas 2008), in which the planetary perturbations and relativistic effect
were included, unlike in Kreutz's (1901) set of elements.

A remarkable feature of the contact-binary model is an outward appearance of
some of its features as chance events that at first sight would not pass the
skeptic's approval.  Because of it, the model may look somewhat vulnerable,
but it is not.  A good example is the case of the daylight comets in AD~363.
The actual amount of information provided by Ammianus Marcellinus may look
rather flimsy, yet it was enough to Seargent (2009) to immediately evoke the
Kreutz sungrazers based solely on the circumstances of observation.  Yet, much
more was involved, as this {\it virtually unique event\/} was highly significant
diag\-nostically for three reasons:\\[0.15cm]
\indent
(i) it was in line with the prediction by a generic model of nontidal
fragmentation that, following the disruption of the progenitor sungrazer
(Aristotle's~comet) at large heliocentric distance, its remains should~at~the
subsequent perihelion appear as a swarm of nearly- simultaneously
arriving fragments;\\[0.15cm]
\indent
(ii) its timing was amazing --- almost exactly midway between the appearances
of Aristotle's comet and the Great Comet of 1106; and\\[0.15cm]
\indent
(iii) it implied an orbital period of about 740~years, consistent with
the period of the Great March Comet of 1843, which itself fits the
pattern!\\[-0.15cm]

Then there was the issue of the {\it missing second predecessor\/} (a ``sibling''
of the Great Comet of 1106) to the two 19th century major Kreutz sungrazers,
which was still missing when I was writing Paper~1!  It was on a hunch that
a novel method was developed in Paper~2, which, against all odds, was successful
in establishing the time of the previous perihelion appearance of comet Ikeya-Seki
(i.e., its parent comet) with a high degree of confidence.  It came as a surprise
that Ikeya-Seki could under no circumstances derive from the 1106 comet.  Rather,
the computed date in late 1139 agreed, with the uncertainty of $\pm$2~years, with
April 1138, the time of the previous appearance of the brightest nucleus of the
Great September Comet of 1882, Ikeya-Seki's more massive twin, which happened to
be determined --- and promptly dismissed as inaccurate --- by Marsden (1967).

As if not enough, this surprise was followed by the detection of a plausible
candidate in Ho's (1962) catalogue, whose estimated perihelion time occurred
at most five months after the date derived by Marsden.  And --- to crown the
achievement --- the sets of orbital elements of the 1882~comet and Ikeya-Seki
integrated back to 1138 agreed much better with one another than when integrated
back to 1115 by Marsden, as seen in Table~4 reproduced from Paper~2.

Could {\it all\/} coincidences in the preceding paragraphs be fortuitous?  Not
in my opinion.

Other points of contention exist between the alternative version of the
two-superfragment model (Sekanina \& Chodas 2007) and the contact-binary
model.  One is the relationship of sungrazers in a cluster.  In the 2007
paper we considered that ``the sungrazers in one cluster are more closely
related to one another than to the sungrazers in the other cluster,'' a
problem resolved by the contact-binary model (Section~3).  To accommodate
the 17th century and earlier clusters, the range of searched arrival times
of the first-generation fragments was broadened in the 2007 paper and the
comets of February 423 and February 467 were regarded as possible Kreutz
sungrazers, referring to Hasegawa \& Nakano (2001) and to England (2002)
as the sources to argue the choice.

It is ironic that in the context of the contact-binary model the two 5th
century comets are no longer Kreutz suspects at a time when the second
object, in AD~467, was suggested to be an attractive candidate, according to
Mart\'{\i}nez et al.\ (2022).  Fortunately, the scope of tidal-fragmentation
scenarios is wide enough that a sungrazer can move in a Kreutz-like orbit
without being a Kreutz sungrazer.  This is particularly true for
a fragment that separated from the progenitor near perihelion {\it
before\/} the Kreutz system's birth.  The comet of 467 could have separated
from Aristotle's comet at perihelion in 372~BC, in which case the orbital
elements of the two objects would have been essentially identical, with the
exception of their periods:\ 735~yr for Aristotle's comet, 838~yr for the
comet of 467.

Having described the issues that were behind the major model overhaul necessitating
the introduction of the contact binary, I now list four features {\it shared\/}
by both the two-superfragment model and the new model.  The first feature
is apparent from the models' names, which imply that fragmentation
began with an event involving {\it two\/} objects:\ the two superfragments
in the former hypothesis versus the two lobes of the contact binary in the
latter.  The second shared feature is the stipulation that the events
responsible for the birth of the fragments took place at {\it very large
heliocentric distance\/}, which implies that the fragments were to pass the
subsequent perihelion nearly simultaneously, in a swarm --- the point already
referred to above.  The third shared feature is the process of cascading
fragmentation, employed in the 2007 investigation.  And the last feature, in
line with the third, is the young age of the Kreutz system, close to two
millennia, the difference between the two models amounting to less than
20~percent.  Both models predict that more brilliant, naked-eye sungrazers
should arrive at perihelion in the coming decades.           

\section{Perihelion Distances of the SOHO\\Dwarf Sungrazers} 
In the context of the contact-binary model, the orbital work on the dwarf
sungrazers detected in the SOHO coronagraphic images focused in Paper~1
almost exclusively on the longitude of the ascending node, especially
the relationship between the nominal values derived by Marsden and the
population-diagnostic true values.  Very little attention was directed
toward the distribution of perihelion distances among the 193 objects of
Table~1.  This shortcoming is being rectified in this section.

The problem with the perihelion distances of the dwarf Kreutz comets is ---
unlike with the nodal longitudes --- the absence of an obvious algorithm for
converting Marsden's nominal values to true values, accounting for the major
nongravitational effects.  The several dwarf sungrazers investigated in detail by
Sekanina \& Kracht (2015) showed that, as a rule, the true perihelion distances
of these objects were closer to the solar radius than their nominal values.
A safe conclusion could be made about those among the comets that were subjected
to relatively minor nongravitational effects and whose nominal nodal longitudes
were close to the true nodal longitudes.  As expected, their true and nominal
perihelion distances differed very little.  In any case, plots of the perihelion
distance as a function of the nodal longitude should provide a fitting line of
attack.

\begin{figure}[t]
\vspace{-1.06cm}
\hspace{2.75cm}
\centerline{
\scalebox{0.77}{
\includegraphics{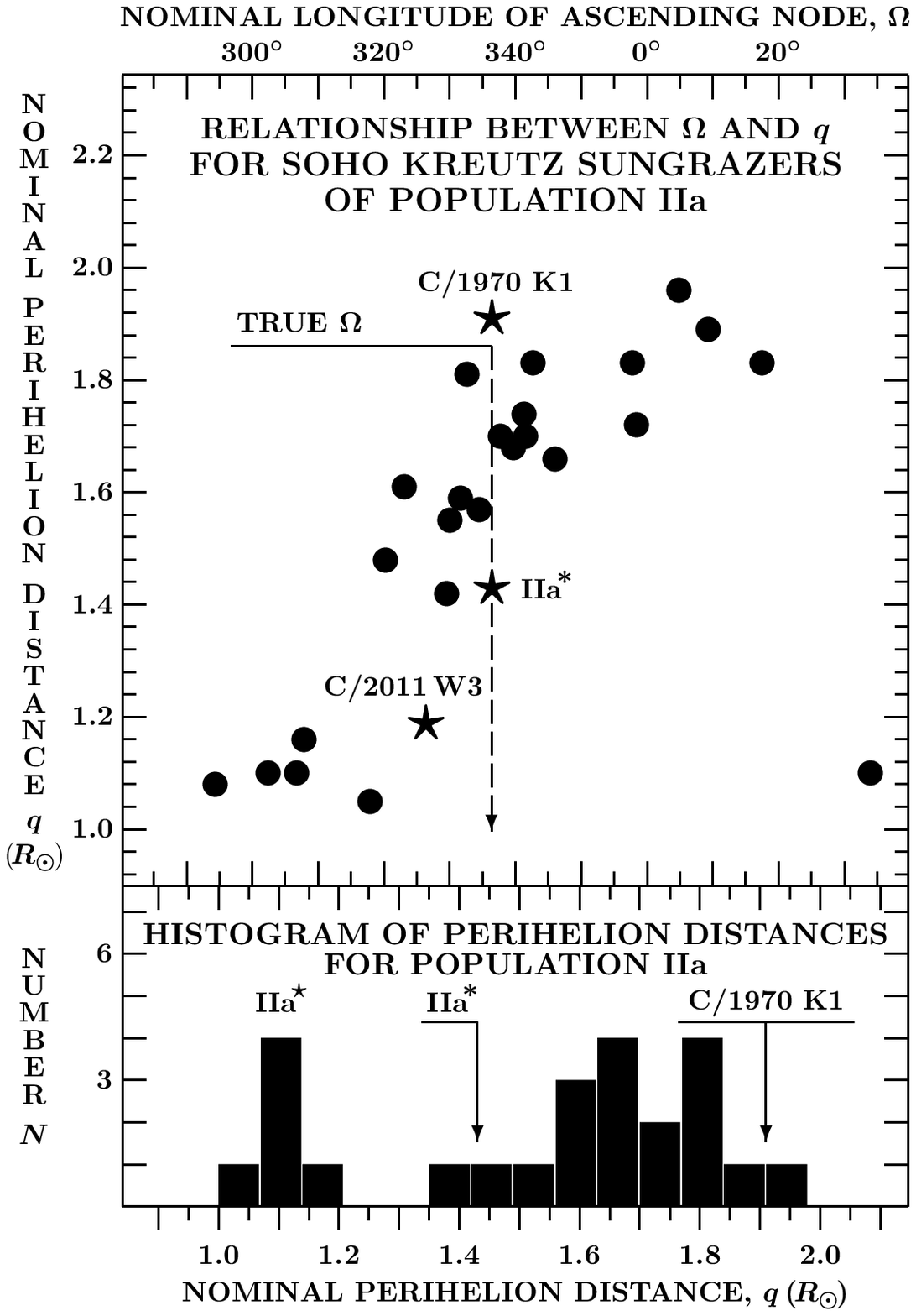}}} 
\vspace{-9.5cm}
\caption{{\it Upper panel\/}:\ Relation between the nominal longitude of the ascending
node, $\Omega$ (equinox J2000), and the nominal perihelion distance, $q$, for 24~dwarf
Kreutz sungrazers of Population~IIa imaged exclusively by the C2 coronagraph of the
SOHO observatory.  The data are sharply split into two parts:\ (i)~a major subset with
\mbox{$q > 1.4\:R_\odot$}, which includes objects in orbits ranging from those similar
to C/1970~K1 to those{\vspace{-0.08cm}} close to a presumed alternative parent
IIa{\large $^\ast$} to comet{\vspace{-0.085cm}} Lovejoy (see Paper~1); and (ii)~a minor
subset with \mbox{$q < 1.2\:R_\odot$}, which is designated as IIa{\large $^\star$}
and contains objects in orbits with perihelion distances very close to that of comet
Lovejoy, perhaps a transition to Population~III.  {\it Lower panel\/}:\ Histogram of
perihelion distances{\vspace{-0.065cm}} of the 24 dwarf sungrazers, showing a sharp
peak of Population~IIa{\large $^\star$} and a broad bulge by the major subset, whose
boundaries are approximately{\vspace{-0.07cm}} delineated by the perihelion distances
of C/1970~K1 and IIa{\large $^{\!\ast}$}. (Adapted from Paper~1.){\vspace{0.6cm}}}
\end{figure}

This method of inquiry was used in the only attempt in Paper~1 to find out
whether an investigation of perihelion distances could contribute meaningfully
to the study of the Kreutz system's structure.  The attempt concerned the
24~members of Population~IIa listed in Table~1 and the conclusion, seen
from Figure~8 adapted from Paper~1, was the clear discrimination of the
perihelion-distance distribution into two distinct subpopulations.  The upper
part is a plot of the nominal perihelion distance as a function of the nominal
nodal longitude, while the lower part is a histogram of the nominal perihelion
distances.  The main set of 18~objects had perihelia between 1.4~{\Rsun}
and 2.0~{\Rsun} and on the plot of the nominal nodal longitude against the
nominal perihelion distance this subpopulation covers a continuous area
that includes comet White-Ortiz-Bolelli.  A subpopulation IIa$^\ast\!$, which
was introduced in Paper~1 to offer an alternative birth scenario for comet
Lovejoy, blends as part of this main set.  The remaining six sungrazers,
referred to in Paper~1 as Population~IIa$^\star\!$, have perihelia between
1~{\Rsun} and 1.2~{\Rsun}, which are in Figure~2 distinguished from the
rest of Population~IIa by special symbols.  The difference between the
perihelion distances of the subpopulation IIa$^\star$ and comet
White-Ortiz-Bolelli, 0.7~{\Rsun} or more,{\vspace{-0.03cm}} implies
a transverse component of the separation velocity in excess of 5~m~s$^{-1}$.
The meaning of this result is unclear, potentially indicating that the
fragmentation process was even more complex.

\begin{figure}[b]
\vspace{-0.3cm}
\hspace{2.75cm}
\centerline{
\scalebox{0.77}{
\includegraphics{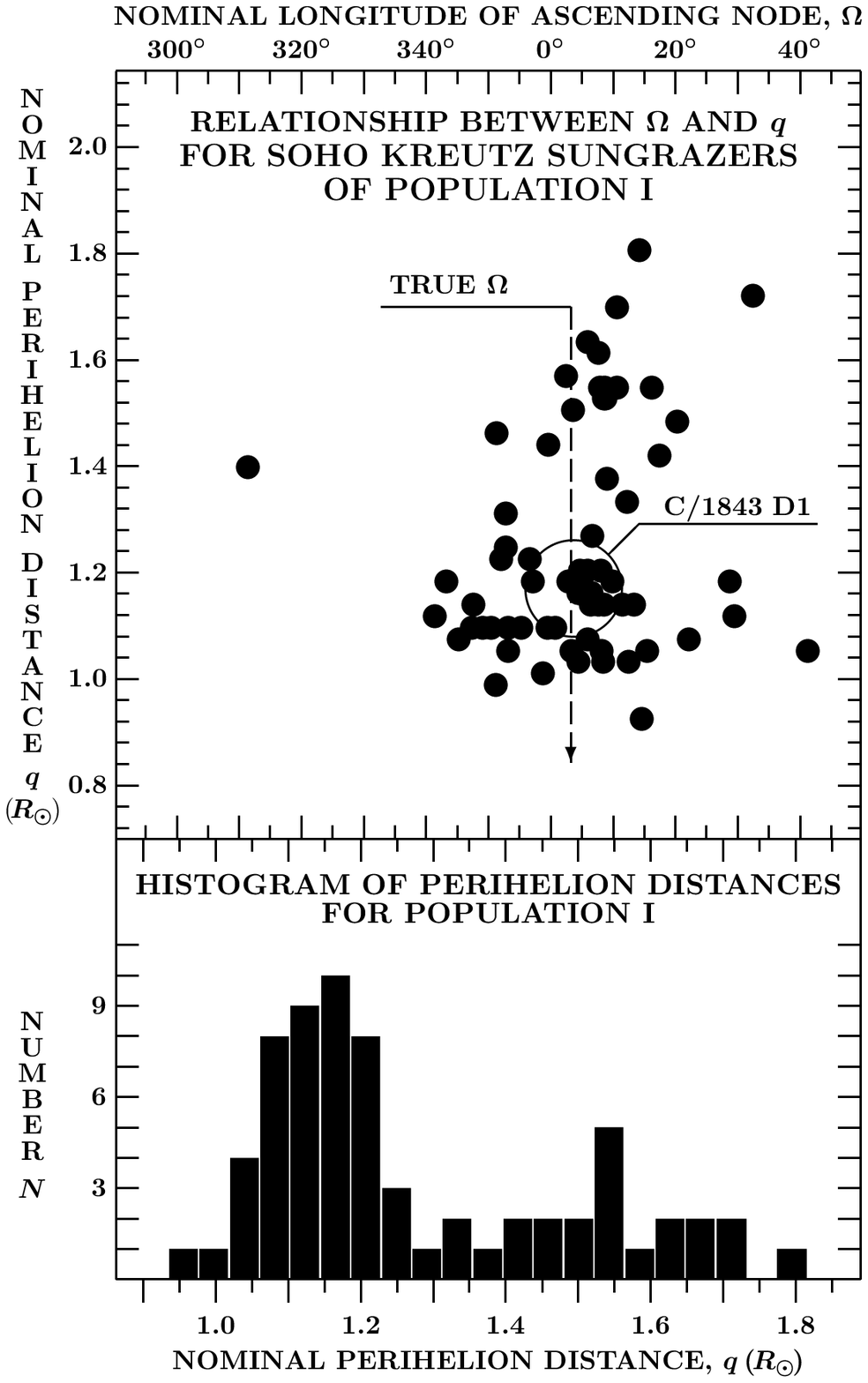}}} 
\vspace{-8.11cm}
\caption{{\it Upper panel\/}:\ Relation between the nominal longitude of the
ascending node, $\Omega$, and the nominal perihelion distance, $q$, for 66
dwarf Kreutz sungrazers of Population~I imaged exclusively by the C2 coronagraph
of the SOHO observatory.  The Great March Comet of 1843 (C/1843~D1) is at the
center of the large open circle.  {\it Lower panel\/}:\ Histogram of nominal
perihelion distances of the 66 dwarf sungrazers.  The main peak coincides with
the position of the 1843 sungrazer, a secondary peak is near \mbox{$q =
1.55$}~{\Rsun}.{\vspace{0.25cm}}}
\end{figure}

It should be pointed out that a general pattern on the $q(\Omega)$ plots for
the Kreutz dwarf sungrazers, of the type displayed in Figure~8, is affected
by the planetary perturbations and subjected to two categories of major
effects:\ (i)~the process of cascading fragmentation, resulting in ever
smaller dimensions of the SOHO objects; and (ii)~the outgassing-driven
nongravitational acceleration, causing the problems with the line
of apsides.  Either effect is capable on its own to lead to
the ultimate disintegration of the objects shortly before perihelion.
   
The fragmentation events, proceeding throughout the orbit, influence the orbital
elements because of the accumulating separation-velocity effects:\ in the nodal
longitude because of the normal component, in the perihelion distance owing to
the transverse component.  The result in the plot of $q(\Omega)$ is a trend in
Figure~8 along a tilted line with an average slope of $dq/d\Omega > 0$, which
is apparent among the data points at \mbox{$q > 1.4$ {\Rsun}}.  On the other
hand, the nongravitational effect in the out-of-plane direction (see Table~4
of Sekanina \& Kracht 2015), affects the angular elements, the nodal line in
particular.  The trend in the $q(\Omega)$ plot should be along a line with
\mbox{$dq/d\Omega \simeq 0$}, which is apparent in the figure among the data
points at \mbox{$q \leq 1.2$ {\Rsun}}.  The gap in the perihelion distance
between 1.2 and 1.4~{\Rsun} could be the product of a major fragmentation event
that has not been accounted for by the present model.

The next task is to examine the $q(\Omega)$ relations and the perihelion-distance
distributions of the other populations.  I employ the same technique as in
Paper~1 and start with the most extensive Population~I.  Its histogram in
Figure~9 resembles that of Population~IIa in Figure~8 in that it also exhibits
two peaks, but differs from it by the enormity of the peak just above the
photosphere, whose position coincides with the perihelion distance 1.17~{\Rsun}
of the Great March Comet of 1843; the relationship between this comet and
Population~I is unquestionable.  A second sharp but much smaller peak is located
near 1.55~{\Rsun}, otherwise the distribution is rather flat.{\pagebreak}

%
\begin{figure*}[t]
\vspace{-6.33cm}
\hspace{1cm}
\centerline{
\scalebox{0.8}{
\includegraphics{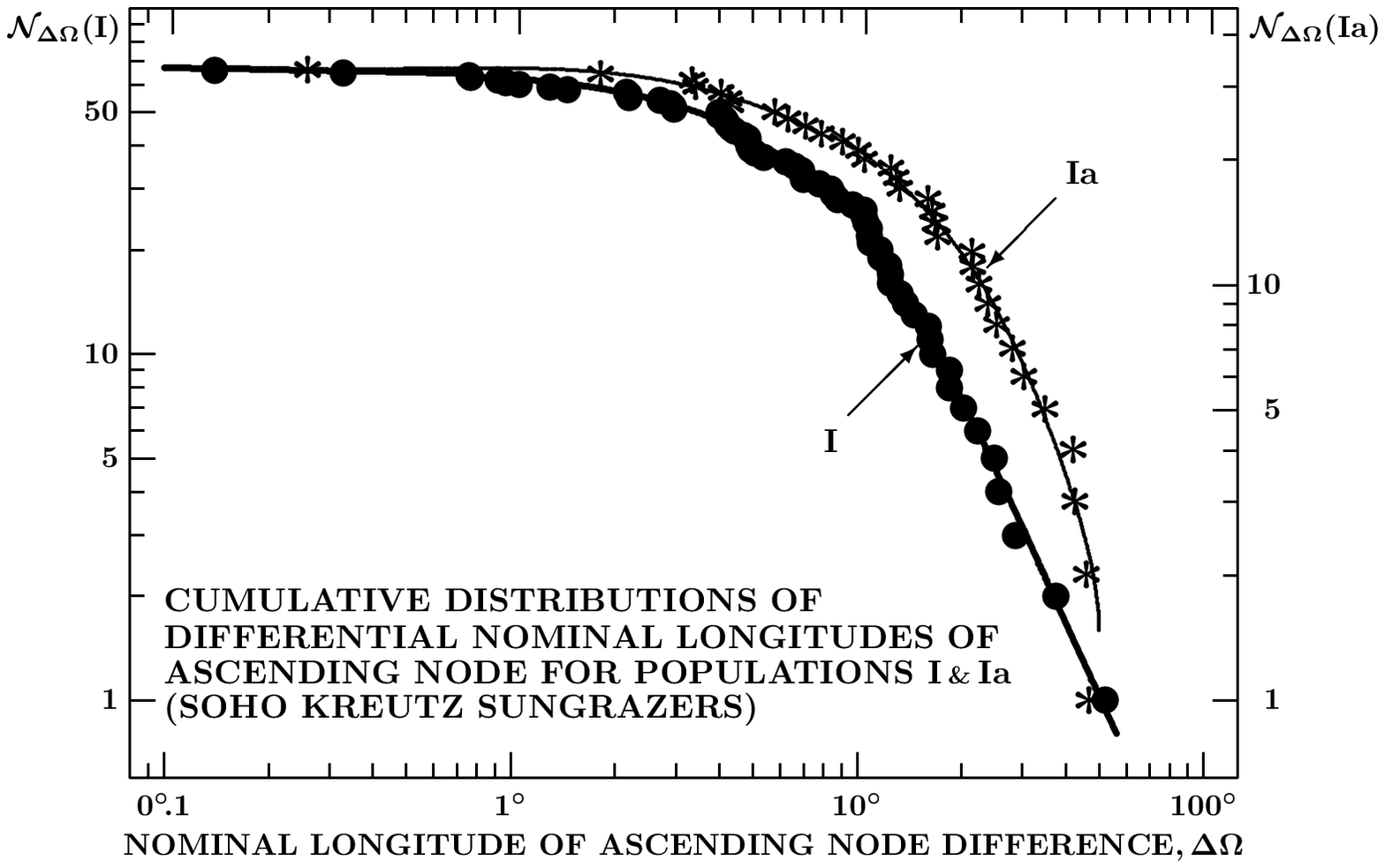}}} 
\vspace{-9.72cm} 
\caption{Cumulative distributions of the differential nominal longitude
of the ascending node from Equation~(12) for the SOHO Kreutz sungrazers of
Populations I (bullets) and Ia (asterisks).  Plotted is the number of comets
whose differential nominal nodal longitudes equal or exceed the plotted limit.
The distribution of the Population~I objects strongly resembles the distribution
of the nongravitational parameter $\cal A$ among the long-period comets (Sekanina
2021b), whereas the distribution of the Population~Ia objects is substantially
steeper at large values of $\Delta \Omega$ and flatter at \mbox{$\Delta \Omega
\rightarrow 0$}.  The difference suggests that different proceesses prevailed
in the two populations.{\vspace{0.8cm}}}
\end{figure*}

Since the difference, $\Delta \Omega$, between the nominal value of the longitude
of the ascending node for a dwarf sungrazer, $\Omega$, and the presumed parent
comet's value, $\Omega_0$,
\begin{equation}
\Delta \Omega = | \Omega - \Omega_0 |, 
\end{equation}
can serve in a first approximation as a proxy parameter that measures the
magnitude of the normal component of the nongravitational acceleration of the
dwarf comet, the cumulative distribution of $\Delta \Omega$ should offer further
information on the process that governs the dispersal of fragments and complement
the results offered by the $q(\Omega)$ relation.  For Population~I the log-log
plot of the cumulative distribution is presented in Figure~10, in which
${\cal N}_{\Delta \Omega}$(I) is the number of the population's dwarf sungrazers,
whose values of $\Delta \Omega$ exceed or equal the plotted limit.  At high
values of $\Delta \Omega$ [or low numbers ${\cal N}_{\Delta \Omega}$(I)] the
curve is a straight line showing that
\begin{equation}
{\cal N}_{\Delta \Omega}({\rm I}) \sim \Delta \Omega^{-2.1}, 
\end{equation}
strongly resembling the distribution of the nongravitational parameters ${\cal A}$
for a set of long-period comets, presented in Figure~2 of Sekanina (2021b) and
displaying at large values of ${\cal A}$ a variation of ${\cal A}^{-1.7}$.  This
similarity strengthens a conclusion that the distribution of the SOHO sungrazers of
Population~I was dominated by sublimation effects, with a lesser contribution
from cascading fragmentation.

The $q(\Omega)$ relation and the histogram of nominal perihelion distances for
Population~Pe are exhibited in Figure~11.  The features resemble those of
Population~I in Figure~9, with the exception of the significantly narrower range
of the nominal nodal longitudes, with 20$^\circ$--25$^\circ$ of the true longitude,
suggesting larger objects.

\begin{figure}[b]
\vspace{-0.3cm}
\hspace{2.75cm}
\centerline{
\scalebox{0.77}{
\includegraphics{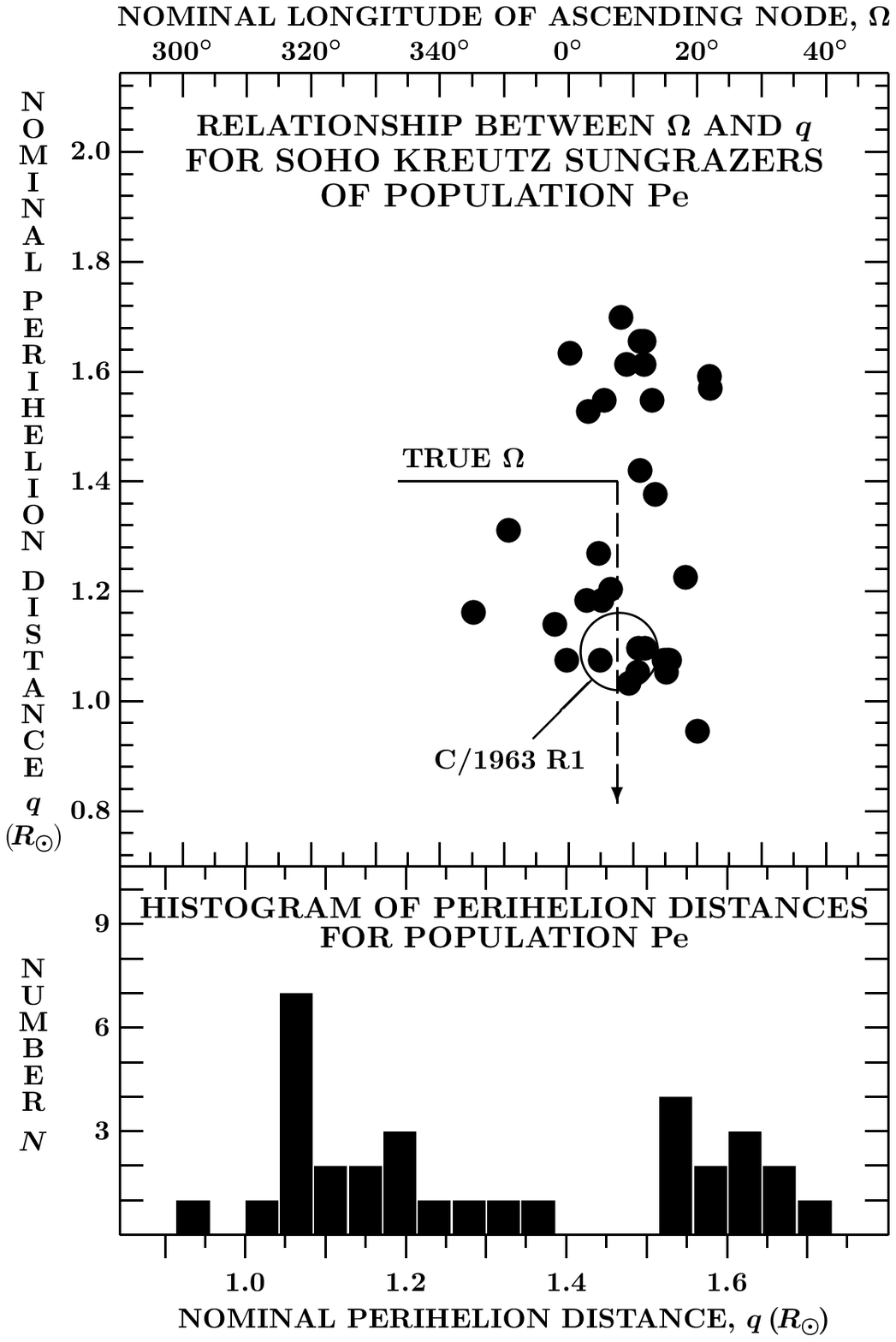}}} 
\vspace{-8.96cm}
\caption{{\it Upper panel\/}:\ Relation between the nominal longitude of the
ascending node, $\Omega$, and the nominal perihelion distance, $q$, for 32 dwarf
Kreutz sungrazers of Population Pe imaged exclusively by the C2 coronagraph
of the SOHO observatory.  Comet Pereyra (C/1963~R1) is at the center of the large
open circle. {\it Lower panel\/}:\ Histogram of nominal perihelion distances
of the 32 dwarf sungrazers.  The main peak is shifted to a slightly lower
perihelion distance, of about 1.07~{\Rsun}, compared to the comet's perihelion
{\vspace{-0.03cm}}distance.  The second peak is near 1.55~{\Rsun}.}
\end{figure}
\begin{figure}[b]
\vspace{-0.5cm}
\hspace{2.75cm}
\centerline{
\scalebox{0.77}{
\includegraphics{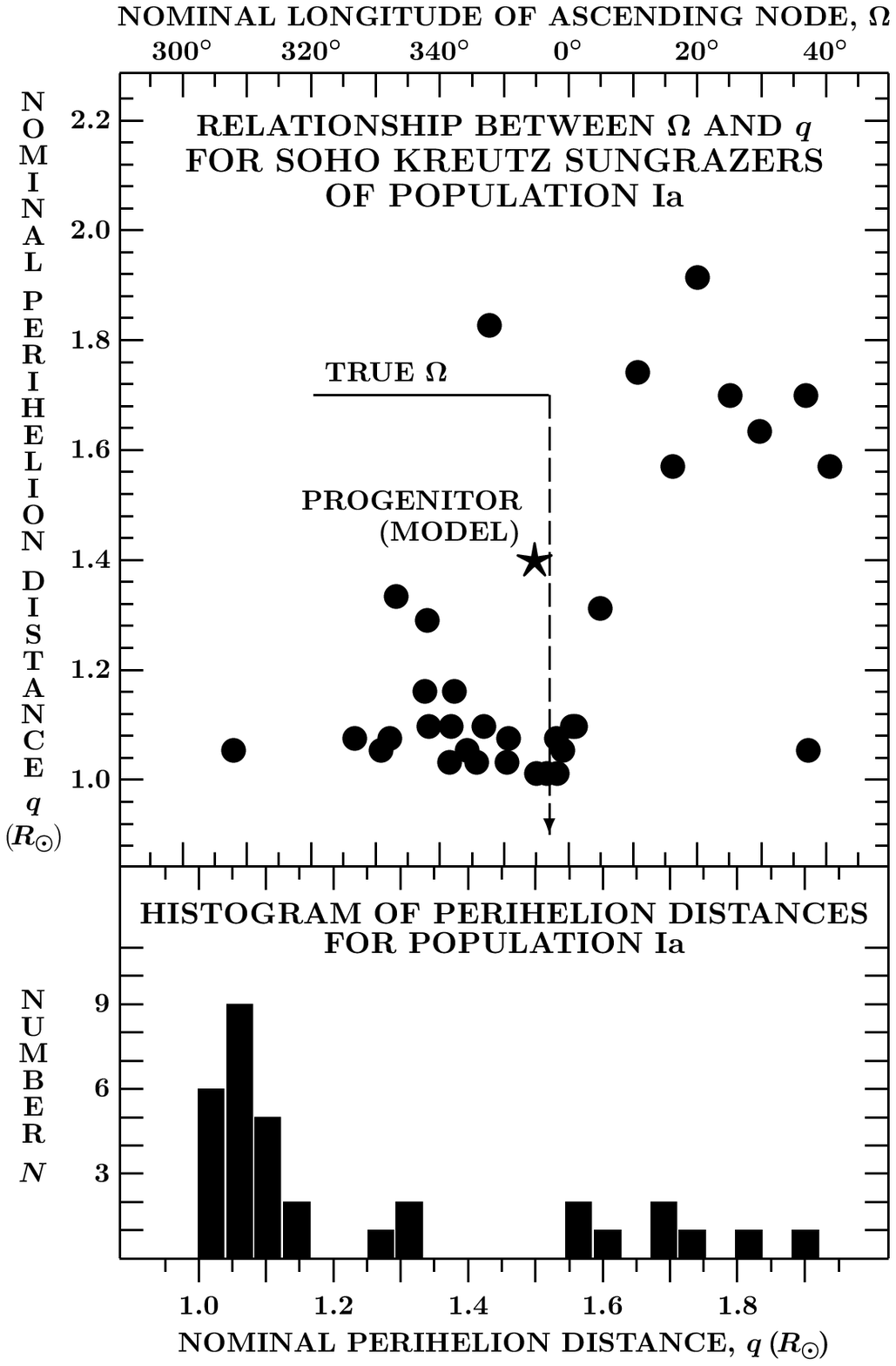}}} 
\vspace{-8.72cm}
\caption{{\it Upper panel\/}:\ Relation between the nominal longitude of the
ascending node, $\Omega$, and the nominal perihelion distance, $q$, for 33 dwarf
Kreutz sungrazers of Population Ia imaged exclusively by the C2 coronagraph
of the SOHO observatory.  The large star is the position of the progenitor
in the working model of Paper~1.  {\it Lower panel\/}:\ Histogram of nominal
perihelion distances of the 33 dwarf sungrazers.  In sharp contrast to Figures~8,
9, and 11, the distribution exhibits only a single prominent peak, located at
a perihelion distance of 1.06~{\Rsun}.  Compared to{\vspace{-0.06cm}}
Population~I, the peak is even nearer the surface of the photosphere and it
is definitely much more narrow.{\vspace{0.09cm}}}
\end{figure}

Figure 12 displays the distribution of nominal perihelion distances among the SOHO
dwarf comets belonging to Population~Ia.  Given the unclear history of this
population, but some connection to the contact binary's neck, the histogram
of nominal perihelion distances is of particular interest.  The results are
stunning, as, contrary to expectation, the distribution displays only a single
peak positioned still closer to the surface of the photosphere than the peak
of Population~I.

To examine more closely the meaning of the unexpected result, the cumulative
distribution of the differential nodal longitude $\Delta \Omega$ is compared
in Figure~10 with the cumulative distribution of Population~I.  The two
distributions are clearly very different.  While Population~I follows at
large $\Delta \Omega$s the relation (13), the curve of Population~Ia is
there considerably steeper, running way above Population~I when normalized.
Figure~13 reveals the nature of the difference between the two distributions:\
replacing the power law by an exponential law suggests that the rate of
growth in the number of fragments varies with their number,
\begin{equation}
d{\cal N}_{\Delta \Omega}({\rm Ia}) = -C_0 \, {\cal N}_{\Delta \Omega}({\rm Ia})
  \, d(\Delta \Omega), 
\end{equation}
or
\begin{equation}
{\cal N}_{\Delta \Omega}({\rm Ia}) = {\cal N}_0({\rm Ia}) \, e^{-C_0
  \Delta \Omega}, 
\end{equation}
where ${\cal N}_0$(Ia) = 33 is the total number of Population~Ia fragments.
From Figure~13, \mbox{$C_0 = 0.058$ deg$^{-1}$}.  The proportionality
between the number of fragments and the rate of growth in their
number is of course an obvious property of colliding particles:\ the more
of them there are, the more they keep running into one another.  Given that
the $\Delta \Omega$ values for Population~Ia were computed with the
progenitor model's value of \mbox{$\Omega_0 = 354^\circ\!$.8} and that the
corresponding value of the model's perihelion distance was 1.40~{\Rsun}, it
is remarkable to see in Figure~12 the fairly crowded area of the nominal
nodal longitudes near $\Omega_0$ on the one hand and a major deficit of
the SOHO sungrazers with the nominal perihelion distances near the model's
perihelion distance on the other hand.  The process of fragmentation appears
to be very efficient in dispersing the fragments' perihelion distances away
from the parent's value.  And again, there appears to be a class of fragments
with their perihelia just above the surface of the photosphere.

Figures 14 and 15 exhibit the relation between the nominal nodal longitude
and the nominal perihelion distance as well as the latter's histogram for,
respectively, Populations~II and III.  Because of the smaller numbers of
sungrazers detected in these populations, they provide less reliable results
than Populations~I and Ia.  However, indications are that once again there
are peaks near 1.1~{\Rsun}, which of course is more surprising for
Population~II.  On the $q(\Omega)$ plot of Figure~15 comet Lovejoy is just
on the outskirts of a clump of SOHO sungrazers.

The three remaining populations --- Pre-I, IIIa, and IV --- have fewer than
ten members each in the collected set and their nominal perihelion distances are
summarized in Table~5.  To the extent that the statistics can at all be
deemed meaningful, the same peak near 1.1~{\Rsun} is again recognized.
RMSD is the root-mean-square deviation.

In summary, the relationships and distributions involving the nominal
perihelion distances of the SOHO dwarf sungrazers appear to suggest that
the process of fragmentation may have been even more complex than is
implied by the contact-binary model.  Overwhelming evidence points to
the detection of excess numbers of the SOHO sungrazers with the nominal
perihelion distances near 1.1~{\Rsun} in apparently every population,
regardless of the perihelion distance of the bright, naked-eye sungrazer,
when known to be associated.  Given that a fragment released from its
parent with a separation velocity, whose transverse component is directed
against the parent's orbital velocity, ends up in an orbit with a smaller
perihelion distance, excessive numbers of the SOHO sungrazers in such
orbits may be surprising only in the populations derived from Lobe~II,
which separated from the progenitor with a positive transverse velocity
(Figure~5).  Since both positive and negative directions are possible, the
perihelion-distance histograms could imply a preferential fragmentation
mode.  On the other hand, one conclusion that the convoluted relationship
between the nominal and true values of perihelion distance allows is that the
amount of information conveyed by the distribution of the nominal perihelion
distances may be rather limited and should not be overrated.
%
%
\begin{figure}
\vspace{-5.17cm}
\hspace{2.46cm}
\centerline{
\scalebox{0.71}{
\includegraphics{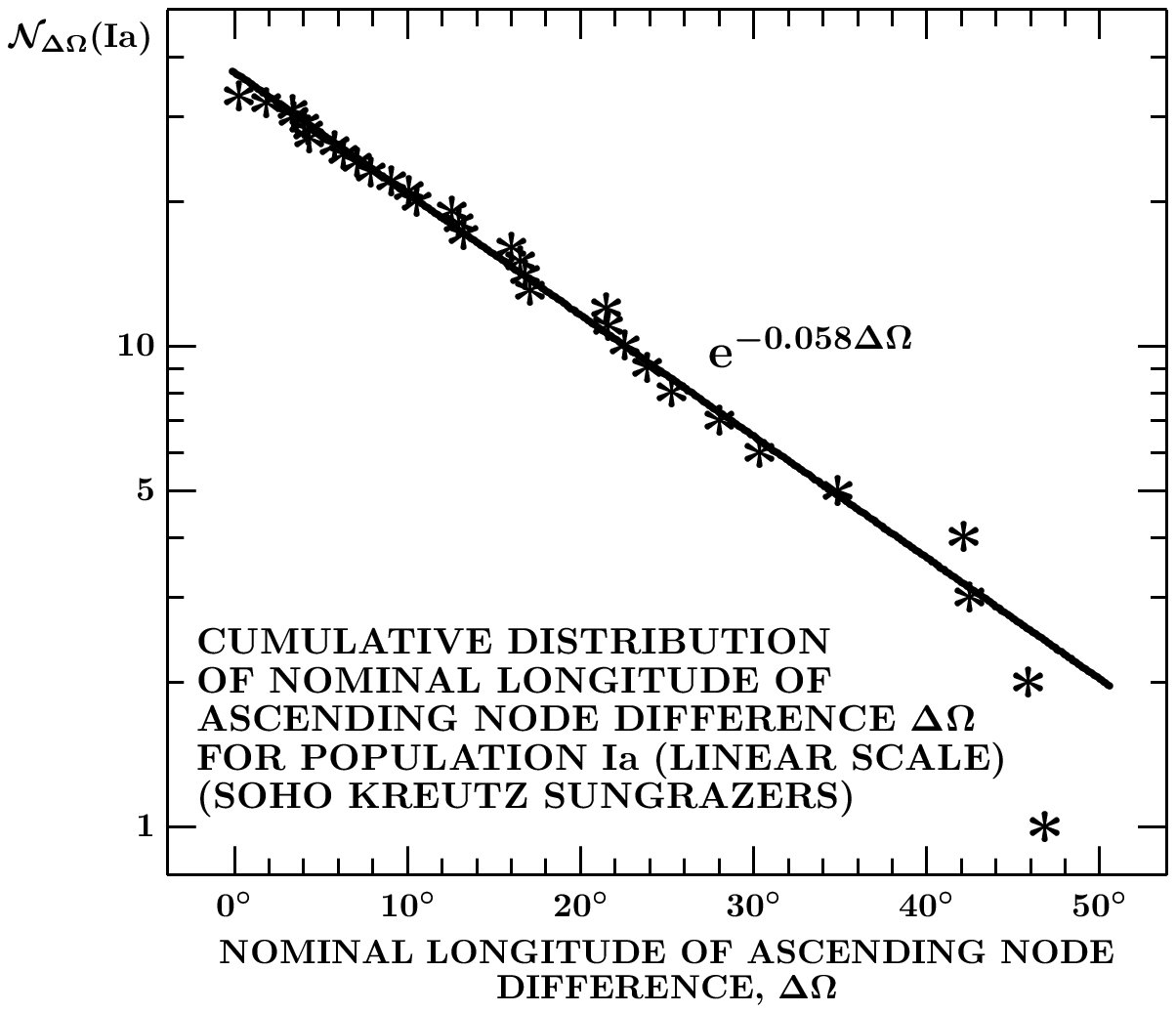}}} 
\vspace{-8.72cm}
\caption{Cumulative distribution of the differential nominal
longitude of the ascending node from Equation~(12) for the SOHO
Kreutz sungrazers of Population~Ia.  Plotted is the number of
comets whose differential nominal nodal longitudes equal or
exceed the plotted limit and follow an exponential law,
apparently suggesting the dominance of fragmentation over
sublimation.{\vspace{0.7cm}}}
\end{figure}

\section{Conclusions} 
This paper has two objectives.  One is to summarize and streamline the
recently introduced contact-binary model for the Kreutz sungrazer system,
whose details were developed and verified by direct orbit integration
in three papers --- 1, 2, and 3 (see the list of references).  This
summary aims primarily at making the presentation of the model more
reader-friendly, but was also necessitated by new information added
stepwise in between the three papers.  A striking example is the derived
birth place of the Great September Comet of 1882 and comet Ikeya-Seki,
an issue that was not tackled until Paper~2, which happened to be
completed last.

%
\begin{figure}
\vspace{-1.03cm}
\hspace{2.75cm}
\centerline{
\scalebox{0.77}{
\includegraphics{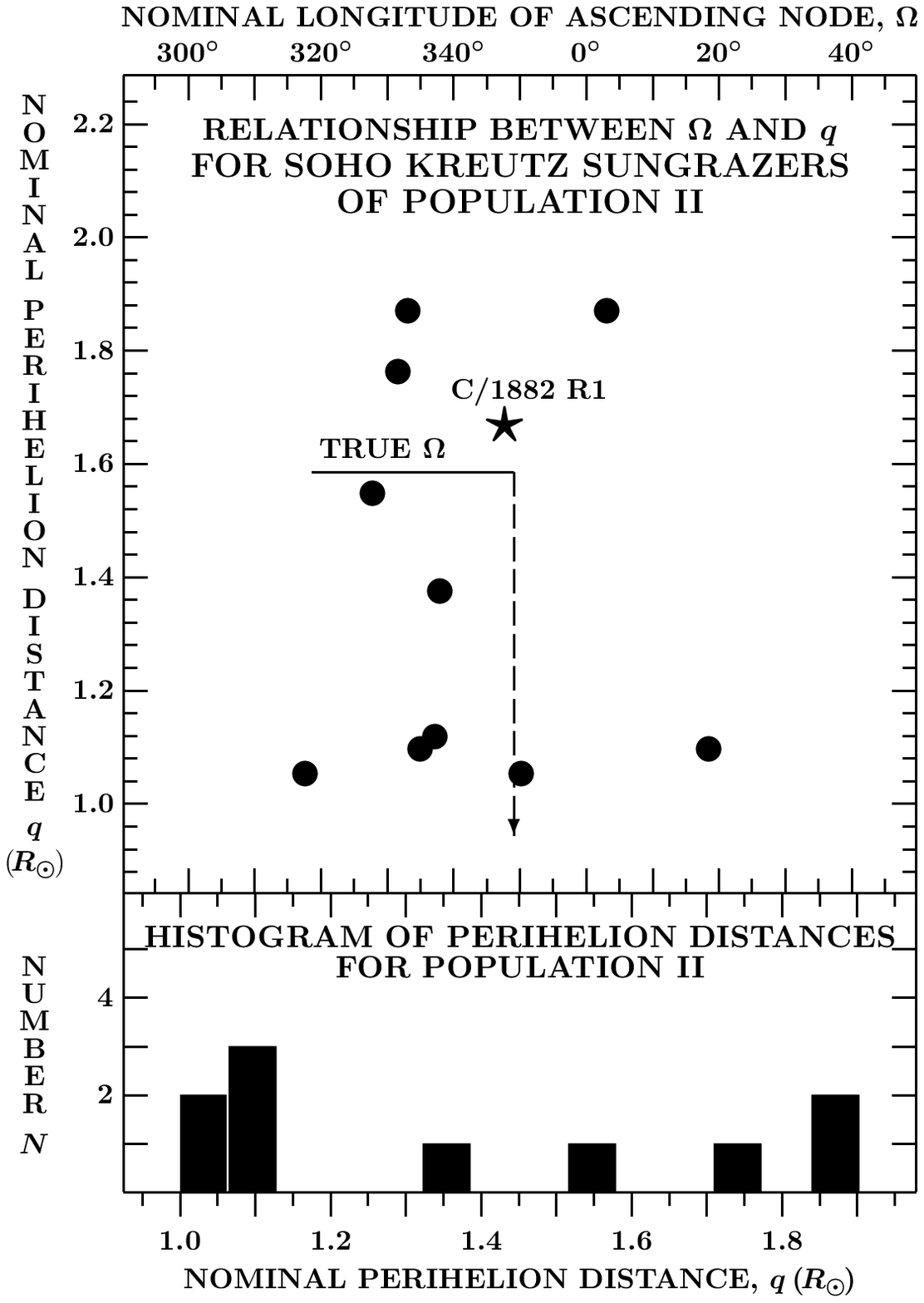}}} 
\vspace{-9.75cm}
\caption{{\it Upper panel\/}:\ Relation between the nominal longitude
of the ascending node, $\Omega$, and the nominal perihelion distance,
$q$, for 10 dwarf Kreutz sungrazers of Population~II imaged exclusively
by the C2 coronagraph of the SOHO observatory.  The large star is the
position of the Great September Comet of 1882.  {\it Lower panel\/}:\
Histogram of nominal perihelion distances of the 10 dwarf sungrazers.
Although the statistics are poor, 50~percent of the data points are
unexpectedly{\vspace{-0.03cm}} below 1.2~{\Rsun} and way below the
perihelion distances of the 1882 sungrazer and Ikeya-Seki.{\vspace{0.8cm}}}
\end{figure}

The contact-binary model begins with Aristotle's comet as the adopted
progenitor of the Kreutz system, presumably a body of formidable size,
perhaps 100~km or more across.  The model assumes that the comet's nucleus
consisted of two masses, referred to as the {\it lobes\/} (and connected by
a narrower neck), which eons ago merged into a single object at a very low
relative velocity.

Whereas the data analysis dictated that the narrative of Papers~1 to 3
proceed from an account of more recent fragmentation episodes, involving
the surviving pieces of the progenitor, to earlier events, involving more
substantial masses, this summary offers an account of the sequence of
fragmentation events in their {\it chronological\/} order, again to the
benefit of the reader.

The initial and most fundamental of these episodes~--- the birth of
the Kreutz system --- was the progenitor's breakup into the two lobes (plus
the neck) near aphelion, at $\sim$160~AU from the Sun.  Perhaps a product
of material fatigue, this event must have occurred at large heliocentric
distance --- an {\it absolutely essential condition\/} ensuring that a
very modest separation velocity should suffice to open up a sizable gap
between the orbits of Lobes~I and II, equaling nearly 20$^\circ$ in
the longitude of the ascending node and about 0.5~{\Rsun} in the
perihelion distance, while keeping the apsidal directions almost
perfectly aligned.  Splitting near the Sun could never explain the
orbital disparity of this magnitude.

%
\begin{figure}
\vspace{-1cm}
\hspace{2.75cm}
\centerline{
\scalebox{0.77}{
\includegraphics{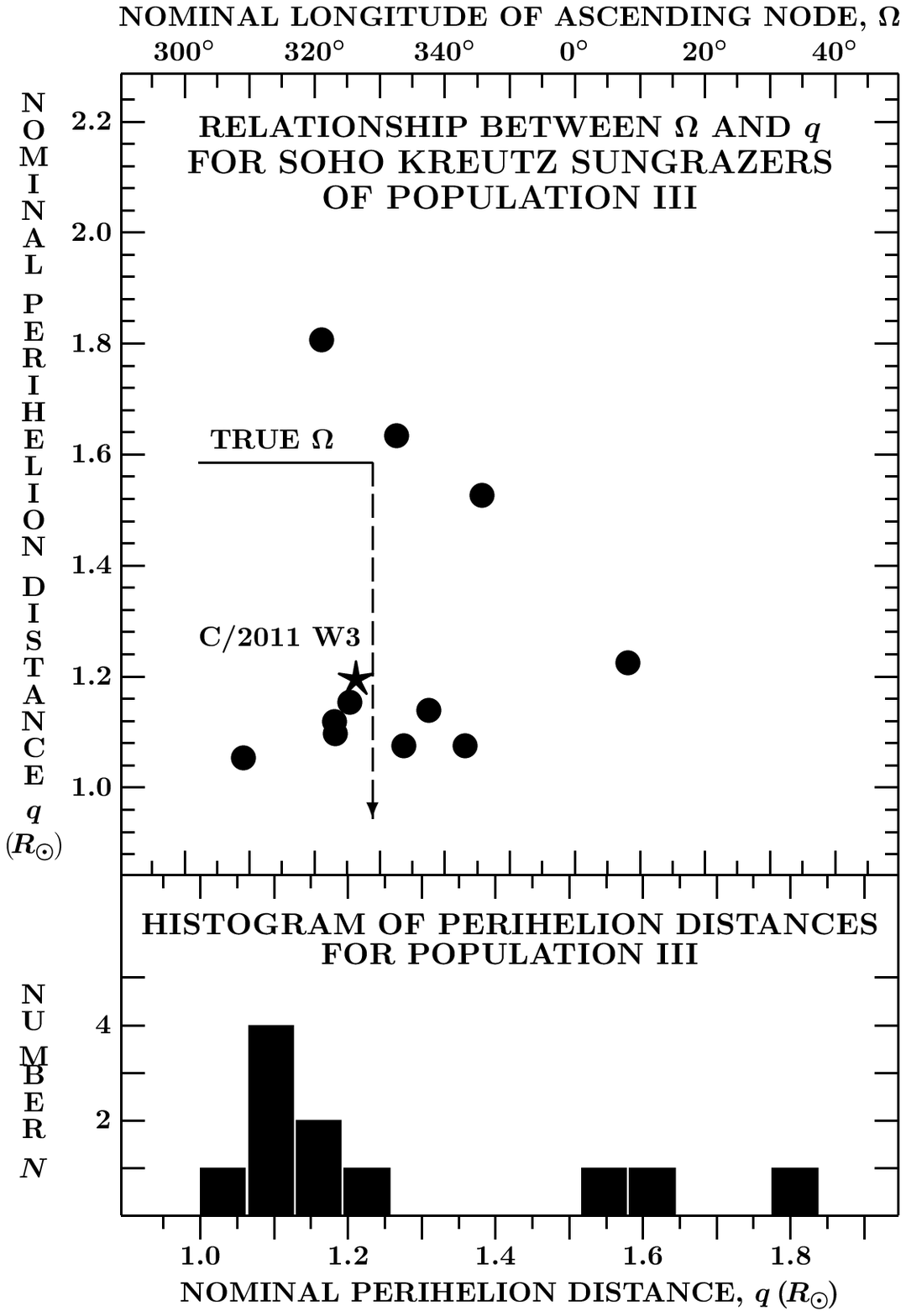}}} 
\vspace{-9.4cm}
\caption{{\it Upper panel\/}:\ Relation between the nominal longitude
of the ascending node, $\Omega$, and the nominal perihelion distance,
$q$, for 11 dwarf Kreutz sungrazers of Population~III imaged exclusively
by the C2 coronagraph of the SOHO observatory.  The large star is the
position of comet Lovejoy.  {\it Lower panel\/}:\ Histogram of nominal
perihelion distances of the 11~dwarf sungrazers.  In spite of the poor
statistics, 70~percent of these distances are smaller than, or comparable
with, the perihelion distance of comet Lovejoy.{\vspace{0.7cm}}}
\end{figure}

The initial breakup was followed, still near aphelion, by events of
secondary fragmentation, as the lobes (especially Lobe~II) continued
to break up, thus generating precursors of the nine populations, whose
existence was established in Paper~1.  The most massive surviving
fragment of Lobe~I did eventually become the Great March Comet of 1843
of Population~I, the most massive fragment of Lobe~II --- the Great
September Comet of 1882 of Population~II.  The smaller fragments of
Lobe~I became the precursors of Populations~Pre-I and Pe, the smaller
fragments of Lobe~II became the precursors of Populations~IIa, III,
IIIa, and IV.  The surviving mass of the neck may have separated from
one of the two lobes (probably Lobe~I) a little later to become the
precursor of Population~Ia.

\begin{table}[t]
\vspace{-4.13cm}
\hspace{5.25cm}
\centerline{
\scalebox{1}{
\includegraphics{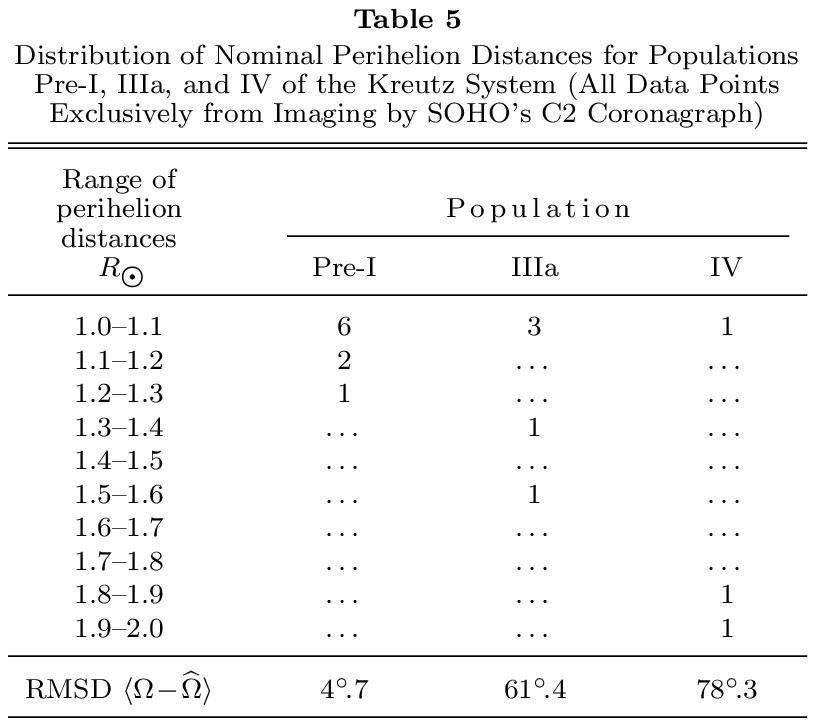}}} TABLE 5
\vspace{-17.75cm}
\end{table}

Because near-aphelion fragmentation of the progenitor and its
lobes was affecting the orbital periods quite insignificantly,
the first-generation fragments were predicted to have arrived
at perihelion in the late 4th century AD nearly simultaneously.
Magnificent in appearance, they must have presented an awesome
sight, consistent with Ammianus' brief, yet fitting comment on the
daylight comets in AD~363 (accompanied by an eloquent and more
elaborate narrative on comets generally).  The inevitable side
effects of the sudden arrival of a swarm of massive comets at
sungrazing perihelion were numerous events of tidal splitting, in
the aftermath of which the nearly identical orbital periods of the
first-generation fragments were abruptly replaced by widely-scattered
orbital periods of the second-generation fragments.  These enormous
changes may have been products of acquired separation velocities, or
of radial shifts in the positions of the centers of mass on the order
of a few kilometers, requiring no separation velocities whatsoever.
Only the most massive of the fragments, whose centers of mass were
moved by very small, subkilometer-sized shifts, ended up in orbits
with periods close to the parent's.  The best example was the
fragment that eventually became the Great March Comet of 1843.
While the first-generation fragment of Population~I had an
orbital period of 735~yr and the population's second-generation
fragment --- the Great Comet of 1106 --- a period of 742~yr,
the 1843 sungrazer itself ended up in an orbit with a period of
737~yr, a remarkably uniform motion from generation to generation.

Because of the large scatter in the orbital periods, the second-generation
fragments were returning to perihelion at very different times.  One can
be more specific only in the cases in which the orbital period of a
third-generation fragment is known with fairly high accuracy, a rather
rare occasion.  Next to Population~I, a solution could be charted for
Population~II, whose second-generation fragment appears to have been the
Chinese comet of 1138, as established in Paper~2 from the data on comet
Ikeya-Seki, a third-generation fragment, and its famous twin, the spectacular
1882 sungrazer.  A different evolutionary path was found for comet Pereyra,
whose likely previous appearance was the comet discovered in September of
1041 rather than the Great Comet of 1106.  The pedigree of comet Lovejoy
could not be determined; for other naked-eye Kreutz sungrazers the history
of orbital evolution is unclear because their orbital periods are unknown.
However, the Great Southern Comet of 1887 is likely to be a fragment of the
Great Southern Comet of 1880, which in turn appears to be a fragment of the
Great Comet of 1106.

On the premise that the mechanism responsible for the dispersal of the
naked-eye Kreutz sungrazers in time and space likewise applied to the
dwarf members seen in the coronagraphic images of the space observatories,
the SOHO spacecraft in particular, one can get some interesting estimates for
the streams of these dwarf objects.  For the currently dominant Population~I,
derived from the Great Comet of 1106, I estimate the total number of dwarf
sungrazers at 300,000 and the duration of the stream (nearly equal to the maximum
orbital period) at $\sim$80,000~yr, even though the rate should be dropping
dramatically with time:\ in 3000~yr from now it should be about 1/10th the
current rate.  Of course by that time the 1843 sungrazer will have passed
perihelion four or five times, presumably resupplying the stream.  One
cannot rule out the possibility that the stream will strengthen with time
until the source is exhausted.  By comparison, the poor showing of the
Population~II stream of dwarf sungrazers is disappointing.        

The second objective of this paper is the examination of one aspect of
the motions of the SOHO dwarf sungrazers that was just barely touched upon
in Papers~1 to 3.  Broadly, this problem concerns their gravitational orbits
derived by Marsden, which, useless when unprocessed, contain important
information that has allowed the detection of the nine populations discriminating
in terms of the longitude of the ascending node.  More specifically, the
unsettled problem involves the nominal perihelion distances, which were
deemed very poorly determined (and published to merely two decimals) by
Marsden, but potentially of a diagnostic value for continuing classification
efforts.  A limited study of this kind in Paper~1 bore only on Population~IIa,
leading to a possible further expansion of the fragmentation process beyond
the presented scenario of the contact-binary model.

The extended investigation of the distribution of the nominal perihelion
distances and their relation to the nominal nodal longitudes for a select
subset of the SOHO dwarf sungrazers, presented in this paper for the first
time, does not unfortunately provide much information about the subject.
The primary result is the acknowledgment of an essentially universal peak
in the histograms of the nominal perihelion distances at about 1.1~{\Rsun},
suggesting a potential preference for fragments released with a negative
transverse component of the separation velocity.  However, given Marsden's
skepticism about the poor quality of his determination of the SOHO sungrazers'
perihelion distances, one cannot rule out that these data may provide no or
very little useful information.

In summary, the contact-binary model offers the picture of a very different
and more complex, though better organized, Kreutz system than was provided
by the two-superfragment model (Sekanina \& Chodas 2004, 2007).  The primary
reason for the refinements is the considerable amount of highly relevant
information~that has been accumulated over the past 15~years, unavailable
when the two-superfragment model was being developed.  Yet it is appropriate
to emphasize that the two models also have features in common that discriminate
them from the earlier models of the Kreutz system:\ the number of initial
fragments limited to two; a very large distance from the Sun at birth;
cascading fragmentation; and young age.  Both models predict that another
cluster of bright Kreutz sungrazers is to arrive at perihelion in the
coming decades.  And we are still waiting for the first naked-eye members
of Populations~Pre-I, Ia,~IIIa,~and~IV.\\

This research was carried out at the Jet Propulsion Laboratory, California
Institute of Technology, under contract with the National Aeronautics and
Space~Administration.\\[-0.2cm]

\begin{center}
{\footnotesize REFERENCES}
\end{center}

\vspace{-0.1cm}
\begin{description}
{\footnotesize
%
%
\item[\hspace{-0.3cm}]
Elkin, W. L. 1883, Astron. Nachr., 104, 281
\\[-0.57cm]
\item[\hspace{-0.3cm}]
England, K. J. 2002, J. Brit. Astron. Assoc., 112, 13
\\[-0.57cm]
%
%
\item[\hspace{-0.3cm}]
Hall, M. 1883, Observatory, 6, 233
\\[-0.57cm]
\item[\hspace{-0.3cm}]
Hasegawa, I., \& Nakano, S. 2001, Publ.\,Astron.\,Soc.\,Japan, 53, 931
\\[-0.57cm]
\item[\hspace{-0.3cm}]
Ho, P.-Y. 1962, Vistas Astron., 5, 127
\\[-0.57cm]
\item[\hspace{-0.3cm}]
Katsonopoulou, D. (ed.) 2017, Helike V:\ Ancient Helike and Aigia-{\linebreak}
 {\hspace*{-0.6cm}}leia.  Athens:\ Helike Society, 310pp
\\[-0.57cm]
%
%
%
\item[\hspace{-0.3cm}]
Kreutz, H. 1891, Publ. Sternw. Kiel, 6
\\[-0.57cm]
\item[\hspace{-0.3cm}]
Kreutz, H. 1901, Astron. Abhandl., 1, 1
\\[-0.57cm]
%
%
%
\item[\hspace{-0.3cm}]
Marsden, B. G. 1967, AJ, 72, 1170
\\[-0.57cm]
\item[\hspace{-0.3cm}]
Marsden, B. G. 1989, AJ, 98, 2306
\\[-0.57cm]
%
%
%
\item[\hspace{-0.3cm}]
Marsden, B. G., \& Williams, G. V. 2008, Catalogue of Cometary{\linebreak}
 {\hspace*{-0.6cm}}Orbits 2008, 17th ed. Cambridge, MA: Minor Planet Center/{\linebreak}
 {\hspace*{-0.6cm}}Central Bureau for Astronomical Telegrams, 195pp
\\[-0.57cm]
%
%
\item[\hspace{-0.3cm}]
Mart\'{\i}nez, M.\,J., Marco, F.\,J., Sicoli, P., \& Gorelli, R. 2022, Icarus,{\linebreak}
 {\hspace*{-0.6cm}}384, 115112
\\[-0.57cm]
\item[\hspace{-0.3cm}]
Matthews, J. 2008, The Roman Empire of Ammianus.  Ann Arbor:{\linebreak}
 {\hspace*{-0.6cm}}Michigan Classical Press, 608pp
\\[-0.57cm]
\item[\hspace{-0.3cm}]
Ramsey, J. T. 2007, J. Hist. Astron., 38, 175
\\[-0.57cm]
\item[\hspace{-0.3cm}]
Rolfe, J. C. 1940, The Roman History of Ammianus Marcellinus,{\linebreak}
 {\hspace*{-0.6cm}}Book XXV. {\tt https://penelope.uchicago.edu/Thayer/E/Roman/}{\linebreak}
 {\hspace*{-0.6cm}}{\tt Texts/Ammian/25$^\ast\!$.html}
\\[-0.57cm]
\item[\hspace{-0.3cm}]
Seargent, D. 2009, The Greatest Comets in History:\ Broom Stars{\linebreak}
 {\hspace*{-0.6cm}}and Celestial Scimitars.  New York:\ Springer Science+Business%
 {\linebreak}
 {\hspace*{-0.6cm}}Media, LLC, 260pp
\\[-0.57cm]
\item[\hspace{-0.3cm}]
Sekanina, Z. 2002, ApJ, 566, 577
\\[-0.57cm]
\item[\hspace{-0.3cm}]
Sekanina, Z. 2021a, eprint arXiv:2109.01297 (Paper 1)
\\[-0.57cm]
\item[\hspace{-0.3cm}]
Sekanina, Z. 2021b, eprint arXiv:2109.11120
\\[-0.57cm]
\item[\hspace{-0.3cm}]
Sekanina, Z. 2022, eprint arXiv:2202.01164 (Paper 3)
\\[-0.57cm]
\item[\hspace{-0.3cm}]
Sekanina, Z., \& Chodas, P. W. 2004, ApJ, 607, 620
\\[-0.57cm]
\item[\hspace{-0.3cm}]
Sekanina, Z., \& Chodas, P. W. 2007, ApJ, 663, 657
\\[-0.57cm]
\item[\hspace{-0.3cm}]
Sekanina, Z., \& Chodas, P. W. 2008, ApJ, 687, 1415
\\[-0.57cm]
\item[\hspace{-0.3cm}]
Sekanina, Z., \& Chodas, P. W. 2012, ApJ, 757, 127 (33pp)
\\[-0.57cm]
\item[\hspace{-0.3cm}]
Sekanina, Z., \& Kracht, R. 2015, ApJ, 801, 135 (19pp)
\\[-0.57cm]
\item[\hspace{-0.3cm}]
Sekanina, Z., \& Kracht, R. 2022, eprint arXiv:2206.10827 (Paper 2)
\\[-0.57cm]
\item[\hspace{-0.3cm}]
Sierks,\,H., Barbieri,\,C., Lamy,\,P.\,L., et al. 2015, Science, 347, a1044
\\[-0.57cm]
\item[\hspace{-0.3cm}]
Stern, S.\,A., Weaver, H.\,A., Spencer, J.\,R., et al. 2019, Science, 364,{\linebreak}
 {\hspace*{-0.6cm}}9771
\\[-0.68cm]
\item[\hspace{-0.3cm}]
Strom, R. 2002, Astron. Astrophys., 387, L17}
%
\vspace{-0.4cm}
\end{description}
\end{document}